\documentclass[%
reprint,
superscriptaddress,
showpacs,
bibnotes,
amsmath,amssymb,
pra,
floatfix,
]{revtex4-1} 

\usepackage{graphicx}
\usepackage{dcolumn}
\usepackage{bm}
\usepackage{multirow}
\usepackage{array}
\usepackage[usenames,dvipsnames]{xcolor}
\definecolor{nblue}{rgb}{0.0, 0.0, 1.0}
\usepackage[colorlinks,linkcolor=nblue,urlcolor=nblue,citecolor=nblue,plainpages=false,pdfpagelabels,breaklinks]{hyperref}
\usepackage{booktabs}
\usepackage{ctable}
\usepackage{upgreek}
\usepackage{mathrsfs}
\usepackage{amssymb}
\usepackage{amsbsy}
\usepackage{color}
\usepackage{cancel}
\usepackage[bbgreekl]{mathbbol}
\usepackage{marginnote}
\usepackage{amsfonts}
\usepackage[caption=false]{subfig}

\newcommand{\bcen}{\begin{center}}
\newcommand{\ecen}{\end{center}}
\newcommand{\btab}{\begin{tabular}}
\newcommand{\etab}{\end{tabular}}
\newcommand{\bdes}{\begin{description}}
\newcommand{\edes}{\end{description}}

\newcommand{\beq}{\begin{equation}}
\newcommand{\eeq}{\end{equation}}
\newcommand{\bea}{\begin{eqnarray}}
\newcommand{\eea}{\end{eqnarray}}

\newcommand{\half}{\frac{1}{2}}
\newcommand{\bary}{\begin{array}}
\newcommand{\eary}{\end{array}}
\newcommand{\benum}{\begin{enumerate}}
\newcommand{\eenum}{\end{enumerate}}
\newcommand{\bitem}{\begin{itemize}}
\newcommand{\eitem}{\end{itemize}}

\newcommand{\eqn}[1] {Eqn.~(\ref{#1})}

\newcommand{\sect}[1] {Sec.~\ref{#1}}

\newcommand{\fig}[1]{fig.~\ref{#1}}
\newcommand{\Fig}[1]{Fig.~\ref{#1}}

\makeatletter

\newcommand{\Rmnum}[1]{\expandafter\@slowromancap\romannumeral #1@}
\makeatother

\newlength{\myfigwidth}
\newlength{\myhalffigwidth}
\newcommand{\dW}{{\mathbb{W}}}

\newcommand{\dU}{{\mathbb{U}}}
\newcommand{\dA}{{\mathbb{A}}}

\newcommand{\db}{{\mathbb{b}}}
\newcommand{\dC}{{\mathbb{C}}}

\newcommand{\dS}{{\mathbb{S}}}
\newcommand{\dO}{{\mathbb{O}}}

\newcommand{\domg}{{\mathbb{\bbomega}}}

\newcommand{\SU}[1]{SU($#1$)}
\newcommand{\eff}[1]{#1_{\textup{\mbox{eff}}}}
\newcommand{\fs}[1]{f_s^{(#1)}}
\newcommand{\fh}[1]{f_h^{(#1)}}

\makeatletter
\newcommand{\thickhline}{%
    \noalign {\ifnum 0=`}\fi \hrule height 2pt
    \futurelet \reserved@a \@xhline
}
\newcolumntype{"}{@{\hskip\tabcolsep\vrule width 2pt\hskip\tabcolsep}}

\newcommand{\mylabel}[1]{\label{#1}} 

\newsavebox{\measurebox}

\usepackage{bbm}

\newcommand{\titlename}{Phases of Attractive Fermi Gases in Synthetic Dimensions}

\usepackage{lineno}

\begin{document}



\title{\titlename}
\author{Sudeep Kumar Ghosh}
\email{sudeep@physics.iisc.ernet.in}
\affiliation{Centre for Condensed Matter Theory, Department of Physics, Indian Institute of Science, Bangalore 560 012, India}
\author{Sebastian Greschner}
\affiliation{Institut f\"{u}r Theoretische Physik, Leibniz Universit\"{a}t Hannover, 30167 Hannover, Germany}
\author{Umesh K. Yadav}
\affiliation{Department of Physics, Lovely Professional University, Phagwara - 144411, Punjab, India}
\author{Tapan Mishra}
\affiliation{Department of Physics, Indian Institute of Technology, Guwahati, Assam - 781039, India}
\author{Matteo Rizzi}
\affiliation{Institut f\"{u}r Physik, Universit\"{a}t Mainz, Staudingerweg 7, D-55099 Mainz, Germany}
\author{Vijay B. Shenoy}
\affiliation{Centre for Condensed Matter Theory, Department of Physics, Indian Institute of Science, Bangalore 560 012, India}


\date{\today}

\begin{abstract}
A novel way to produce quantum Hall ribbons in a cold atomic system is to use $M$ hyperfine states of atoms in a $1$D optical lattice to mimic an additional ``synthetic dimension''. A notable aspect here is that the \SU{M} symmetric interaction between atoms manifests as ``infinite ranged'' along the synthetic dimension. We study the many body physics of fermions with attractive interactions in this system. We use a combination of analytical field theoretic and numerical density matrix renormalization group (DMRG) methods to reveal the rich ground state phase diagram of the system, including  novel phases such as squished baryon fluids. Remarkably, changing the parameters entails unusual crossovers and transitions, e.~g., we show that  increasing the magnetic field (that produces the Hall effect) may convert a ``ferrometallic'' state  at low fields to a ``squished baryon superfluid''(with algebraic pairing correlations) at high fields. We also show that this system provides a unique opportunity to study quantum phase separation in a multiflavor ultracold fermionic system.
\end{abstract}


\maketitle


Cold atomic systems have emerged as one of the most promising platforms to realize low dimensional quantum systems that have continued to provide many fascinating phenomena both in equilibrium and nonequilibrium~\cite{Bloch2008,Giorgini2008,Guan2013}.
Recent advances of introducing synthetic gauge fields~\cite{Dalibard2011,Lin2009B,Cheuk2012} and realizing synthetic dimensions~\cite{Celi2014,Stuhl2015,Mancini15} have opened a new direction of research. ``Synthetic dimension'' (SD)~\cite{Celi2014} is created by coherently coupling $M$ hyperfine (HF) states of atoms loaded in a $1$D optical lattice via Raman lasers to generate an additional spatial dimension. It is described by the Hamiltonian
\bea\mylabel{eqn:KE}
\!\!\!\!\!\!\!H_0 = -t \sum_{i;\gamma=1}^M  C^\dagger_{i+1,\gamma}C_{i,\gamma}  + \sum_{i;\gamma = 1}^{M-1} \Omega_\gamma^i C^\dagger_{i,\gamma+1} C_{i,\gamma} + \mbox{h.c.} \;,
\eea
where, $C^\dagger_{i,\gamma}$ and $C_{i,\gamma}$ are the fermion operators associated with site $i$ of the optical lattice (coordinate $x_i = i d$, $d$ is the lattice spacing) and HF state $\gamma$. The hopping amplitude from one site to its nearest neighbor is $t$. $\Omega_\gamma^i = \Omega_\gamma e^{-ik_l x_i}$ ($k_l$ is the wave vector of the Raman laser) is the Raman coupling of the HF state $\gamma$ with $(\gamma+1)$ and its phase results in an effective magnetic flux $\phi = k_l d$ per plaquette of the synthetic lattice. By an appropriate choice of $\phi$, Hofstadter model~\cite{Jain2007} in a finite Hall ribbon of width $M$ is realized. Beautiful recent experiments~\cite{Stuhl2015,Mancini15} have demonstrated this proposal.

A remarkable feature of atoms with $M$ HF states used in this scheme is that their contact interaction, 
\beq\mylabel{eqn:HU}
H_U = - \frac{U}{2} \sum_{i,\gamma,\gamma'} C^\dagger_{i,\gamma} C^\dagger_{i, \gamma'} C_{i,\gamma'}C_{i,\gamma} \,\,,
\eeq
of strength $U$ (here, we consider attractive interaction $U>0$) is \SU{M}  symmetric~\cite{Chin06,Fukuhara07,Taie12,DeSalvo10,Gorshkov10,Cazalilla14,
Pagano14}. In the absence of the Raman couplings ($\Omega_\gamma = 0$), this problem has been well studied~\cite{Klingschat10,Capponi08,Rapp07,Pohlmann13,Guan2013}. The ground state is a fluid of $M$-body bound states of fermions (\SU{M} singlets) that are dubbed as ``baryons'', in analogy with similar $M$-body \SU{M} singlets (e.~g. \SU{3} proton) arising in high energy physics. With Raman couplings, the system is a Hall ribbon with infinite ranged interaction along the synthetic dimension. The key open question is: what are the many body phases of fermions in this system? The goal of this paper is to address this outstanding issue. We uncover the rich phase diagram of the system with attractive interactions~\cite{repnote} using a combination of field theoretic and numerical DMRG methods. Interestingly, it supports a variety of phases including ferrometallic fluids, squished baryon~\cite{Sudeep2015} fluids and more. There are also intriguing crossovers/transitions that are quite unusual: a ferrometal (generalized spin polarized Fermi fluid) at very small magnetic flux ($\phi \approx 0$) is converted to a squished baryon superfluid (algebraically correlated quasi-condensate of nonlocal pairs or baryons) by increasing the magnetic field ($\phi = \pi$)! Furthermore, we observe a regime of macroscopic quantum phase separation~\cite{Shenoy06} pointing to the possibilities of the SD system to address a variety of issues in condensed matter.

\noindent
{\bf Model Hamiltonian:} The physics of the SD system is most transparently viewed in a different basis (``flavor'' basis labeled by $\zeta$). The fermion operators in this basis $\db_i \equiv \{b_{i,\zeta}; \zeta = 1, \ldots, M\}^T$ are unitarily~\cite{Grass14,Sudeep2015} related to $\dC_i \equiv \{C_{i,\gamma}; \gamma = 1, \ldots, M\}^T$ by $\dC_i = \dU_i \db_i$ where $\dU_i$ is a unitary matrix. In this basis, the total Hamiltonian ${\cal H} = H_0 + H_U$ of the system can be recast as
\beq
\mylabel{eqn:Ham}
{\cal H} =  -t \sum_i \left(\db^\dagger_{i+1} \dA \db_i + \mbox{h.~c.} \right) + \sum_i \db^\dagger_i \domg \db_i + H_U \;,
\eeq
where, $H_U$ is same as that in \eqn{eqn:HU} with $C_{i \gamma}$ replaced by $b_{i \zeta}$. The SD system is then reduced to a system of $M$ component fermions experiencing a non-Abelian \SU{M} gauge field (encoded in the matrix $\dA = {\dU^{\dagger}_{i+1}}\dU_i$,  which, interestingly, is independent of $i$) and an \SU{M} ``Zeeman field'' $\domg = \mbox{Diag}\{\omega_\zeta; \zeta = 1, \ldots, M\}$ (a diagonal matrix with eigenvalues $\omega_\zeta$). The \SU{M} gauge field produces a flavor orbital coupling, i.~e., the quantum number $\zeta$ can be altered by hopping from site $i$ to its neighbor and is characterized by the non-diagonal elements of $\dA$. Here, we consider $\Omega_\gamma = \Omega$ \cite{Omgnote} and for $M=2$, $H_0$ is schematically depicted in \fig{fig:schematic}. Explicit forms of different matrices are given in Sec.~S$1$ of the supplementary material (SM)~\cite{suppmat}.  

Note that for an arbitrary $\phi$, each $\zeta$ flavor is not individually conserved and the system does not have global \SU{M} symmetry; only a U($1$) symmetry corresponding to the overall particle number conservation survives. We analyze this model using exact numerical methods such as DMRG~\cite{White92,White93,Scholl05} and field theoretic analytical methods~\cite{Gogolin99,Giamarchi04,Miranda03} applied to an effective Hamiltonian (for $\phi=\pi$) that we construct.

\begin{figure}
{
\centering
\includegraphics[width=1.1\myfigwidth]{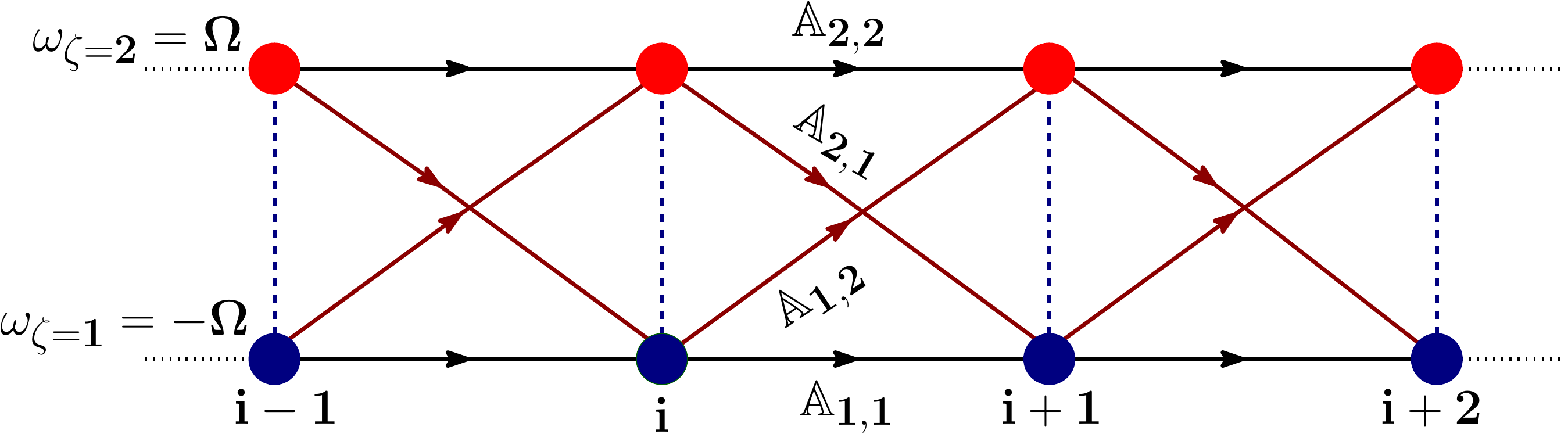}
}
\caption{(Color online) {\bf Kinetic Energy:} Schematic plot of the $M = 2$ SD system with $SU$($2$) gauge and Zeeman field. $\dA_{\zeta,\zeta}$ are the flavor preserving hoppings and $\dA_{1,2}$ and $\dA_{2,1}$ are the flavor-orbital couplings. Note that the $\zeta$ states at any site $i$ are Zeeman split and there is no hopping between them.}
\mylabel{fig:schematic}
\end{figure}

\noindent
{\bf Analytical:} For any $M$, the physics of the system is governed by the dimensionless parameters $t/U$, $\Omega/U$ and $\phi$ (setting $U$ as the basic scale). We note that in the limit of $t=0$, there is an important energy scale in the system, $\Omega_c = \frac{M-1}{4 \cos\left({\frac{\pi}{M+1}}\right)} U$ that determines the Raman coupling strength at which an $M$-baryon is destabilized by the Zeeman field. Throughout this paper, we focus on density $n \lesssim 1/M$ ($n=N/(LM)$ is the average number of particles per site of the synthetic lattice having $L$ optical lattice sites and $N$ particles).

Analytical results are possible for $t/U \lesssim 1$ in the following cases:

\noindent
{\bf i) \underline{$\Omega \ll  \Omega_c$, $\phi=0$}:} Here, the ground state is made of $M$-baryons~\cite{Klingschat10,Capponi08,Rapp07,Pohlmann13} which form a quasi-superfluid state described by an effective theory having central charge $c=1$. This is analogous to a Luther-Emery (LE) phase~\cite{Luther1974} with gapless ``charge'' excitations (which make up the unit central charge) and gapped flavor excitations. Within canonical ensemble, $M$-particle excitations are of lowest energy. 

\noindent
{\bf ii) \underline{$\Omega \gg \Omega_c$, $\phi=0$}:} Such a system has populated states only in the $\zeta = 1$ manifold as the Zeeman energy dominates over all the other energy scales. The system is a nearly free gas of fermions and corresponds to a flavor polarized ``ferrometallic'' state with lowest $1$-particle excitations. E.g. for $M=2$ case, the state is made of particles with HF states polarized in the ``$x$'' direction, i.~e. the magnetic polarization ${\cal P} \equiv \sum_{i,\gamma,\gamma'} \langle C^{\dagger}_{i,\gamma} \tau^{x}_{\gamma,\gamma'} C_{i,\gamma'} \rangle \neq 0$, where $\tau^x$ is the first Pauli matrix.

\noindent
{\bf iii) \underline{$\Omega \ll \Omega_c$, $\phi=\pi$}:} This case is same as the case~{\bf i)} since here $\Omega_\zeta \approx 0$ implies no effect of $\phi$.

\noindent
{\bf iv) \underline{$\Omega \gtrsim \Omega_c$, $\phi=\pi$}:} This particular case has an effective description in terms of ``spin-$\half$'' fermions due to its special hopping structure and an emergent U($1$)$\times$U($1$) symmetry (see Sec.~S$2$ of SM~\cite{suppmat}). Field theoretic description of the state suggests that it becomes a LE liquid (comprising of nonlocal ``spin-$\half$'' singlets) with dominant pair correlations having unit central charge (spin is gapped) and lowest $2$-particle excitations (see Sec.~S$3$ of SM~\cite{suppmat}). Also, it has ${\cal P}=0$ for $M=2$.

These analytical results allow us to arrive at some startling conclusions. Consider $\Omega \gg \Omega_c$; then, starting from a ferrometallic state at $\phi = 0$, we can obtain a state with leading pair correlations by increasing $\phi$ to $\pi$. For any $M$, increasing the magnetic field through the plaquettes thus transforms a ferrometal to a quasi-superfluid of squished pairs!

\begin{figure}[!b]
{
\centering
\includegraphics[width=1.12\myfigwidth]{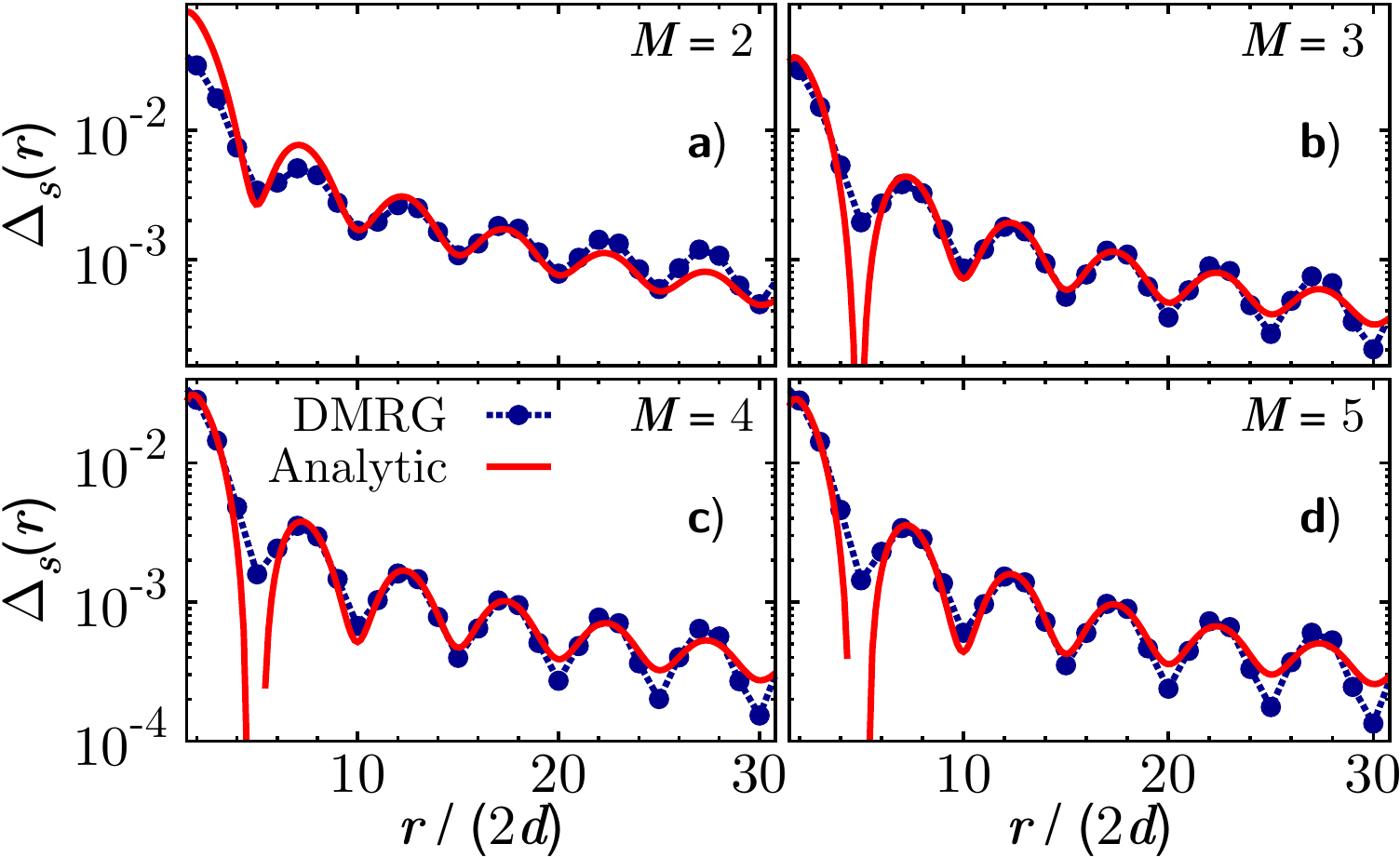}
}
\caption{(Color online) {\bf Nonlocal pair correlation:} Comparison of DMRG and field theoretic results of the algebraic decay of the nonlocal pair correlation $\Delta_s(r)$ for $\pi$-flux with $t/U=0.5$, $\Omega/U = 4$, $L=160$ and $N=32$. Panels a), b), c) and d) show the $M = 2$, $3$, $4$ and $5$ cases respectively.}
\mylabel{fig:bosonization}
\end{figure}

\noindent
{\bf Numerical results:} These intriguing results clearly motivate a detailed numerical study to understand this crossover along with other possibilities which are not immediately evident from an analytical treatment. The next sections discuss numerical DMRG~\footnote{We perform large scale finite system DMRG simulations for system sizes upto $L=160$ with open boundary conditions keeping upto $1000$ matrix states.} results of the manybody ground state phases. 

We characterize different phases by several observables. First, we look for convenient local measures that provide signatures of the nature of the baryons (squished or otherwise) that make up the ground state. Defining the operators ${\hat{O}_{i, h}}^{(p) \dagger} = b_{i, 1}^{\dagger} \ldots b_{i, p}^{\dagger}$ and ${\hat{O}_{i, s}}^{(p) \dagger} = b_{i, 1}^{\dagger} b_{i+1, 1}^{\dagger} \ldots b_{i+p-1, 1}^{\dagger}$, their average local correlations are called $\fh{p} = 1/L \sum_{i}\langle {\hat{O}_{i, h}}^{(p) \dagger} \hat{O}_{i, h}^{(p)} \rangle$ and $\fs{p} = 1/L \sum_{i}\langle {\hat{O}_{i, s}}^{(p) \dagger} \hat{O}_{i, s}^{(p)} \rangle $ respectively, where $p$ can take values $2$, $3$, $\ldots$, $M$. A state with the usual $M$-body baryon will have a dominant value of $\fh{M}$ together with a vanishing $\fs{M}$ but the opposite provides signature of squished baryons. We also look at a nonlocal pair correlation function, $\Delta_s(r) = \langle {\hat{O}_{i, s}}^{(2) \dagger} {\hat{O}_{i+r, s}}^{(2)} \rangle$ with $r$ even, which captures the squishing effect. In addition, we compute the lowest $m$-particle excitation energy $\Delta E_m$ and the von-Neumann entanglement entropy $S_{\mathrm{vN}}$ for a subsystem of length $l$ and extract the central charge ($c$) by fitting $S_{\mathrm{vN}}$ to the Calabrese-Cardy (CC) formula~\cite{Calabrese09} (see Sec.~S$7$ of SM~\cite{suppmat}).

\noindent
{\bf Analytics vs.~DMRG for $\phi=\pi$:} We begin the discussion of the numerical results by comparing them with the analytical results of the effective field theory of the limiting case {\bf iv)} described earlier. In \fig{fig:bosonization}, we show the comparison of $\Delta_s(r)$ for different $M$ SD systems. As is evident, the numerical results are in excellent agreement with those predicted by the field theory, confirming the analytical prediction of the ``squished baryon'' quasi-condensate. It is also noted that $\Delta_s(r)$ decay algebraically and is indeed the dominant correlation of the system for this case (see Sec.~S$3$ of SM~\cite{suppmat}). 

\begin{figure}[t]
{
\centering
\includegraphics[width=1.1\myfigwidth]{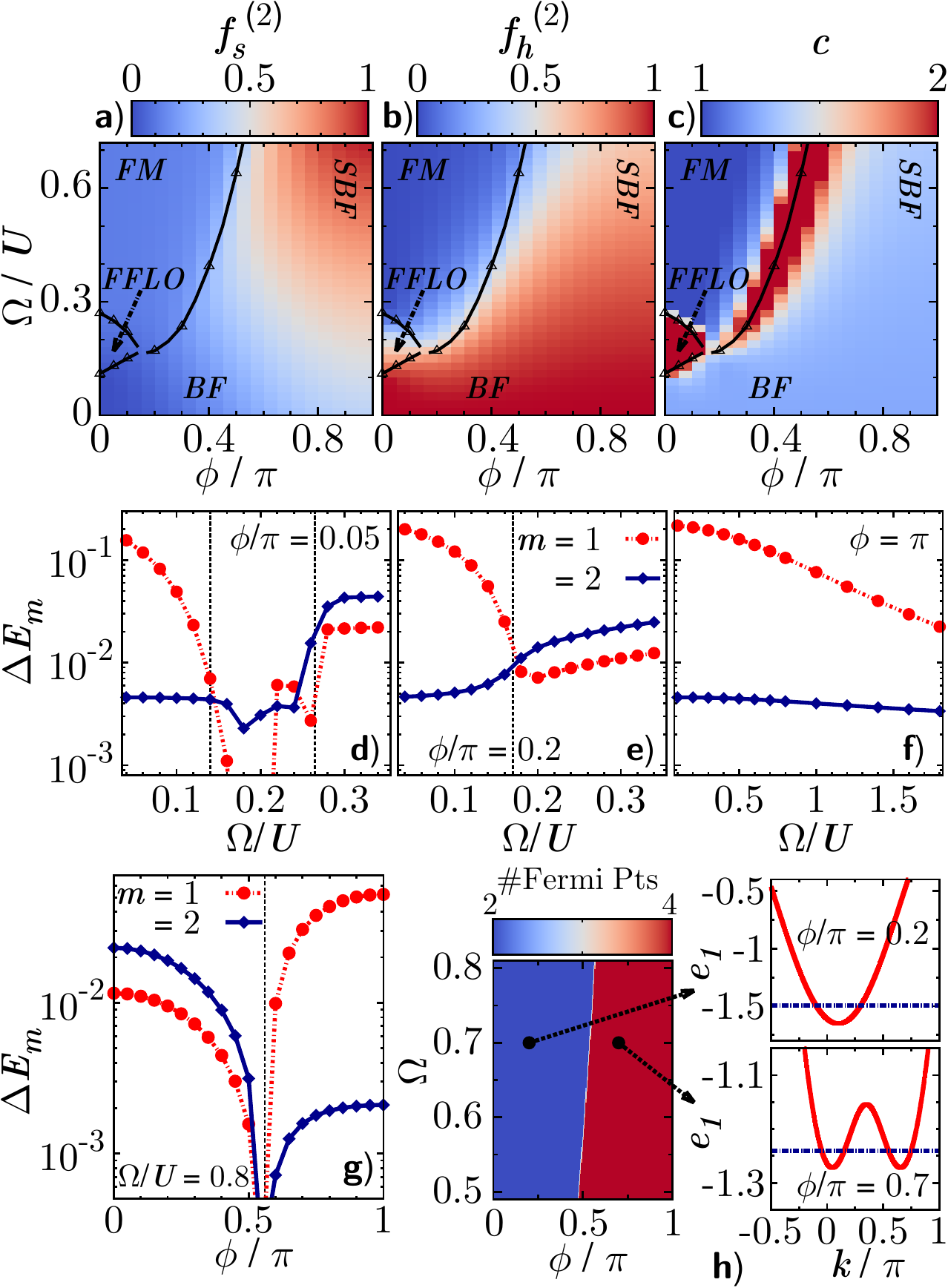}
}
\caption{(Color online) {\bf $M=2$ Results:} $t/U=0.5$, $n=0.1$ and $L=80$. Phase diagrams in the $\Omega$-$\phi$ plane corresponding to normalized~\cite{Note3} $\fs{2}$ and $\fh{2}$, and the central charge $c$ are shown in a), b) and c) respectively. Symbols depict estimates of transition points by DMRG (see Sec. S$6$ of SM~\cite{suppmat}) and the solid lines are guides to the eye. Excitation spectra are shown, along different cut directions of the phase diagrams, first in d), e) and f) versus $\Omega$ for $\phi/\pi = 0.05$, $0.2$ and $1$ respectively and then in g) versus $\phi$ for fixed $\Omega/U=0.8$. Dashed lines show the transition points between different phases seen in the phase diagrams. In the large $\Omega > \Omega_c$ limit, variation of the number of Fermi points of corresponding non-interacting system in the $\Omega$-$\phi$ plane is shown in the left panel of h). Its right panels show the structure of the Fermi surface with the blue lines showing the chemical potential in the first band ($e_1(k)$) for the chosen filling.}
\mylabel{fig:M2}
\end{figure}

\noindent
{\bf DMRG Results ($M=2$):} \Fig{fig:M2} shows the DMRG results for the $M = 2$ system~\footnote{Also, see other studies~\cite{Carr2006,Orignac2001,Petrescu2013,Petrescu2015} in similar two leg ladders with repulsive interaction.} as a function of $\Omega/U$ and $\phi$. Starting the discussion at $\phi\approx0$, we see that for small $\Omega$, $\Omega \ll \Omega_c \leq U/2$ for finite $t$, $\fs{2}$ is small while $\fh{2}$ is large~\footnote{Normalized by the maximum value in the $\Omega$-$\phi$ plane shown in the phase diagram.} indicating the presence of the usual $2$-baryon -- this is the $2$-baryon quasi-superfluid ($2$-BF) (see \fig{fig:M2}(\textbf{a} and \textbf{b})). The signature of LE physics is further corroborated by the unit value of the central charge  (\fig{fig:M2}\textbf{c}), and the lowest excitation being a $2$-particle one (\fig{fig:M2}\textbf{d}). For $\phi \approx 0$ and $\Omega \gg \Omega_c$, we see that there are no pairing correlations $\fs{2},\fh{2} \approx 0$; $c=1$ and lowest $1$-particle excitation, all confirming the expected ``free'' gas of $\zeta=1$ fermions -- the ferrometallic (FM) state. These two phases are separated by a Fulde-Ferrell-Larkin-Ovchnikov (FFLO) phase~\cite{Feiguin2011} at an ``intermediate'' value of $\Omega$. Moving to an ``intermediate'' value of $\phi = 0.2 \pi$, we see that BF and FM phases occur in the expected regimes (\fig{fig:M2}\textbf{e}) without any intervening FFLO phase and a rather sharp transition occurs at $\Omega/U \approx 0.17$. Turning to the case of $\phi=\pi$, we find the usual $2$-BF state for $\Omega \ll \Omega_c$, which smoothly crosses over to the squished baryon fluid (SBF) at large $\Omega$ (\fig{fig:M2}\textbf{f}) -- throughout this process the central charge remains unity and the lowest excitations are 2-particle ones (confirming the analytical prediction). 

 The most remarkable aspect of the phase diagram occurs at $\Omega > \Omega_c$. In this limit, it is evident from \fig{fig:M2}\textbf{g} that with changing $\phi$ the FM state at $\phi=0$ switches over to the SBF state at $\phi= \pi$ precisely as anticipated from the analytical theory. This change is intervened by a critical point at $\phi = \phi_t$ (e.g., $\phi_t \approx 0.56 \pi$ for $\Omega/U=0.8$). We argue that a change in the topology of the Fermi surface, i.~e. a Lifshitz transition (LT)~\cite{Lifshitz30,Blanter94}, of non-interacting fermions underlies this critical point as illustrated in \fig{fig:M2}\textbf{h}. The number of Fermi points is shown in the left panel of \fig{fig:M2}\textbf{h} obtained by a construction shown in the right panels of \fig{fig:M2}\textbf{h}. It is now clear that the locus of transition points from FM to SBF can be understood as that of the LTs occurring in the non-interacting system dressed by interactions. Interestingly, a detailed numerical analysis indicates that in this large $\Omega$ limit, there is a finite parity order in the SBF phase (see Sec.~S$7$ of SM~\cite{suppmat}).

\begin{figure}
{
\centering
\includegraphics[width=1.1\myfigwidth]{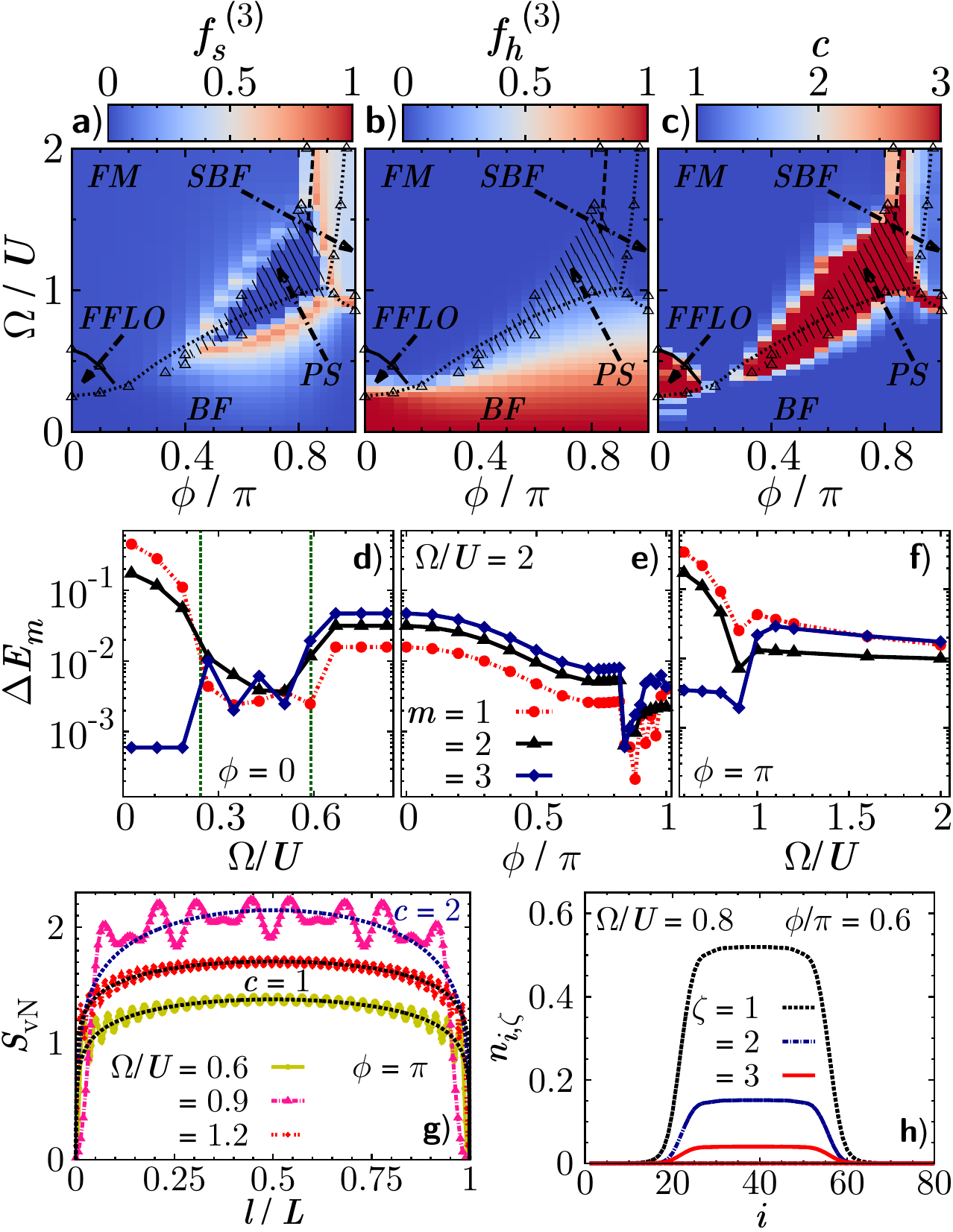}
}
\caption{(Color online) {\bf $M=3$ Results:} $t/U=0.5$ and $n=0.1$. a), b) and c) show phase diagrams corresponding to normalized $\fs{3}$~\cite{mynote}, normalized $\fh{3}$~\cite{Note3} and $c$ respectively ($L=40$). Symbols -- estimates of transition points by DMRG, solid lines -- guides to the eye, dotted lines -- numerical estimates of bound state transitions in dilute limit and dashed lines -- LTs (see Sec. S$5$ and Sec. S$6$ of SM~\cite{suppmat}). d), e) and f) show excitation spectra along different cuts of the phase diagrams ($L=80$, green dotted lines -- extent of the FFLO region). $S_{\mathrm{vN}}$ of a subsystem of size $l$ with fits to the CC formula~\cite{Calabrese09} is shown in g) for $L=160$ and $\phi = \pi$~\cite{Note4}. h) shows variations of onsite populations $n_{i,\zeta}$ in the PS regime ($L=80$) and clustering of particles near the central site is seen.}
\mylabel{fig:M3}
\end{figure}

\noindent
{\bf DMRG Results ($M=3$):} The main results for $M=3$ are shown in \fig{fig:M3}. For $\phi \approx 0$, the state is a fluid (BF) of 3-baryons (which are fermions) whose presence is indicated by a large $\fh{3}$ along with a vanishing $\fs{3}$ as shown in \fig{fig:M3}(\textbf{a} and \textbf{b}). In fact, this state with $c=1$ prevails for all $\phi$ when $\Omega \ll \Omega_c \leq U/\sqrt{2}$ shown in \fig{fig:M3}\textbf{c}. For $\phi \approx 0$, we obtain the FM state when $\Omega \gg \Omega_c$ and it is separated from the BF state by an FFLO phase (see \fig{fig:M3}\textbf{d}). And, for $\Omega \gg \Omega_c$, from the FM state the system becomes an SBF state made of squished $2$-baryons predicted by the analytical theory having the excitation spectra shown in \fig{fig:M3}\textbf{e} as a function of $\phi$. Also, in the limit $\Omega \gtrsim \Omega_c$, there are interesting LTs occurring in the non-interacting Fermi surface (see Sec.~S$5$ of SM~\cite{suppmat}). The physics at $\phi=\pi$, for this $M=3$ case, has some more interesting aspects than that of $M=2$. Increasing $\Omega$ from $\Omega \ll \Omega_c$ results in an interesting transition from the $3$-BF to SBF state having  $3$-particle and $2$-particle lowest energy excitations respectively (\fig{fig:M3}\textbf{f}). \Fig{fig:M3}\textbf{g}~\footnote{As explained in Sec.~S$7$ of SM~\cite{suppmat}, there are strong finite size effects in the estimation of central charge. To reduce this effect, we consider a larger system size $L=160$ for this case using the enhanced symmetries special to this case.} shows that both of these states have $c=1$ while there is an intermediate regime where the $3$-baryons are destabilized in favor of the squished $2$-baryons resulting in $c=2$. The fact that $c=2$ in this regime suggests that the low energy physics has $2$ types of excitations and constructing a field theoretic description of the state will be an interesting future direction. The picture just described is further confirmed by looking at the populations $n_{i,\zeta}$ of different flavors (see Sec.~S$4$ of SM~\cite{suppmat}).

Matters take a dramatic turn near $\Omega \approx \Omega_c$ and $0.4 \pi \le \phi \le 0.7 \pi$ -- the hatched region in \fig{fig:M3}(\textbf{a}-\textbf{c}). In this regime, the system displays phase separation (PS) as is seen from \fig{fig:M3}\textbf{h} -- the central region is of the highest density. The propensity of PS can be further understood by the increased ``flatness'' of single-particle bands in this regime (see Sec.~S$8$ of SM~\cite{suppmat}). It is indeed noteworthy that this system can be used to study the physics of PS in a multiflavor fermionic system.

{\bf Outlook:} SD systems having upto $M \leq 6$ can be realized using the recently studied Yb$^{173}$ system~\cite{Mancini15} and orbital Feshbach resonance~\cite{Zhang2015,Hofer2015,Pagano2015} in this system can produce the \SU{M} symmetric interaction. Also, the \SU{3} symmetric $M=3$ SD system can be realized using nuclear spin-$1$ Li$^{6}$ atoms with a large negative triplet scattering length~\cite{Bartenstein05}. Different phases in the many body phase diagram can be characterized by using familiar band mapping techniques~\cite{Dahan1996,Kohl2005,Bloch2008,Taruell12} along with photoassociation spectroscopy~\cite{Kinoshita05}. Also, the chiral currents, measured in the experiments~\cite{Mancini15,Stuhl2015}, point towards interesting additional structures of different phases and further characterize them as described in Sec.~S$9$ of SM~\cite{suppmat}. In conclusion, we hope that the novel results presented here will stimulate further experimental works aiming at realizing different phases and their transitions in the SD system. 

\paragraph*{{\bf Acknowledgment:}}
S.~G. acknowledges support by QUEST-LFS (Center for Quantum Engineering and Space-Time Research) and DFG Research Training Group (Graduiertenkolleg) $1729$. S.~G. also thanks Temo Vekua, Luis Santos and Leonardo Mazza for enlightening discussions and acknowledges Leibniz University of Hanover, Germany for computing resources. T.~M. would like to acknowledge the support by the start-up research grant from the Indian Institute of Technology, Guwahati, India. V.~B.~S. is grateful to DST, India and DAE, India (SRC grant) for generous support.

\bibliography{refsqdmrg}

\begin{thebibliography}{88}
\bibitem [{myn()}]{mynote}%
  \BibitemOpen
  \href@noop {} {\emph {\bibinfo {title} {{\rm Normalized by the maximum value
  in the $\Omega$-$\phi$ plane after the phase separation region shown by the
  hatched lines in the phase diagrams is ``masked'' by setting $f^{(3)}_s$ to
  be zero.}}}}\BibitemShut {Stop}%
\end{thebibliography}


\newwrite\tempfile
\immediate\openout\tempfile=junkSM.\jobname
\immediate\write\tempfile{\noexpand{\thepage} }
\immediate\closeout\tempfile

\clearpage

\newpage

\appendix

\renewcommand{\appendixname}{}
\renewcommand{\thesection}{{S\arabic{section}}}
\renewcommand{\theequation}{\thesection.\arabic{equation}}
\renewcommand{\thefigure}{S.\arabic{figure}}

\setcounter{page}{1}
\setcounter{figure}{0}

\widetext

\centerline{\bf Supplemental Material}
\centerline{\bf for}
\centerline{\bf \titlename}
\centerline{}
\centerline{by Sudeep Kumar Ghosh, Sebastian Greschner, Umesh K. Yadav, Tapan Mishra, Matteo Rizzi and Vijay B.~Shenoy}

\centerline{}
Abstract: In this supplementary material, we provide explicit expressions for different matrices used in the Hamiltonian of the SD system, discuss in detail an effective theory description of the $\pi$-flux case in the large $\Omega$ limit and present its field theoretic bosonization results. Furthermore, we discuss variations of populations of different flavors, different Lifshitz transitions occurring in the corresponding non-interacting system and properties of bound states in some limiting cases. Detailed description of our numerical analysis with explicit expressions for different observables under consideration is given. Finally, we describe the physics of phase separation and different aspects of experimentally observable chiral currents. 


\section{Explicit forms of different matrices used in the Hamiltonian}
\mylabel{sec:Amat}
As discussed in the main text, the fermion operators in the flavor basis $\db_i$ are related to the original operators in the HF state basis $\dC_i$ via the unitary transformation $\dC_i = \dU_i \db_i$. Here, the position dependent unitary matrix $\dU_i$ is a product of two matrices given by $\dU_i = \dW_i \dS$. The first one $\dW_i = \mbox{Diag}\{e^{i k^{\gamma}_l x_i}; \gamma = 1, \ldots, M\}$ with $k^{\gamma}_l = (\gamma - 1) k_l$  is a diagonal matrix and it contains the position dependence of $\dU_i$. Whereas the second one $S$ is another unitary matrix and is position independent. Hence, the $\dA$ matrix defined as $\dA = {\dU^{\dagger}_{i+1}}\dU_i$ is naturally position independent and can be further simplified to be
\beq
\dA = \dS^{\dagger} \dO \dS \;,
\eeq 
where, the position independent diagonal matrix $\dO = \mbox{Diag}\{e^{i \phi (\gamma -1)};\gamma = 1, \ldots, M\}$.

Now, for the $\Omega_\gamma = \Omega$ case under consideration, the elements of the $\dS$ matrix are given by
\beq
S_{\gamma,\zeta} = \sqrt{\frac{2}{M+1}} \sin\left( \frac{\pi}{M+1} (M-\gamma+1)(M-\zeta+1)\right)
\eeq
and the elements of the $\domg$ matrix are given by
\beq
\omega_\zeta = -2\Omega\cos\left(\frac{\pi\zeta}{M+1}\right) \;.
\eeq
We note that $\dS$ is a real symmetric matrix for this case and this in turn implies that $\dA$ is a complex symmetric matrix. In particular, for $M = 2$,
\beq
\dA = e^{i\frac{\phi}{2}} \left( \begin{array}{ccccc}
\cos(\frac{\phi}{2}) & i \sin(\frac{\phi}{2}) \\
i \sin(\frac{\phi}{2}) & \cos(\frac{\phi}{2})
\end{array} \right) 
\;\;\;\text{and} \;\;\;\; {\domg} = \mbox{Diag}\{-\Omega, \Omega\} \;,
\eeq
 and for $M = 3$
\beq
\dA =  \left( \begin{array}{ccccc}
e^{i\phi} \cos^2({\frac{\phi}{2}}) & \frac{e^{2 i \phi}-1}{2 \sqrt{2}} & - e^{i \phi} \sin^2(\frac{\phi}{2}) \\
 \frac{e^{2 i \phi}-1}{2 \sqrt{2}} & e^{i \phi} \cos(\phi) & \frac{e^{2 i \phi}-1}{2 \sqrt{2}} \\
 - e^{i \phi} \sin^2(\frac{\phi}{2}) & \frac{e^{2 i \phi}-1}{2 \sqrt{2}} & e^{i \phi} \cos^2\frac{\phi}{2} 
\end{array} \right)
\;\;\;\text{and} \;\;\;\; {\domg} = \mbox{Diag}\{-\sqrt{2} \Omega, 0, \sqrt{2} \Omega\} \;.
\eeq

\section{Effective theory in the large $\Omega$ limit for $\pi$-flux}
\mylabel{sec:EffTheory}

\begin{figure}[!ht]
{
\centering
\includegraphics[width=1.2\myfigwidth]{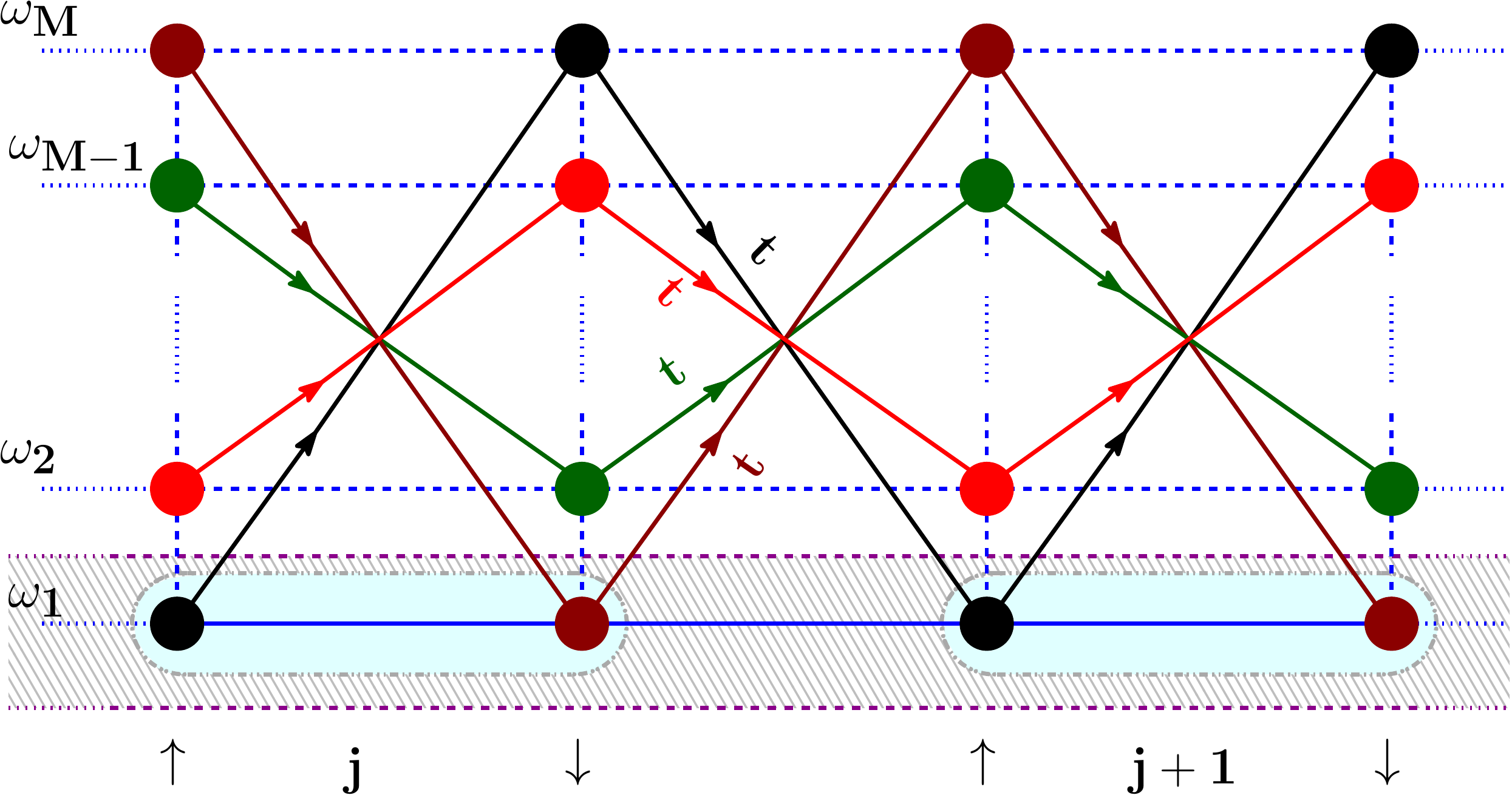}
}
\caption{(Color online) {\bf Kinetic Energy at $\phi=\pi$:} Schematic plot showing the crisscross hopping structure special to the $\phi=\pi$ case. For $\Omega \gg \Omega_c$, low energy sector is marked by the hatched box. In this limit, an effective theory can be constructed in the low energy sector by considering two neighboring sites (shown by cyan boxes) to form a unit cell where the states on the odd (even) physical sites can be thought of as spin $\uparrow$ (spin $\downarrow$) states. The onsite energies of different $\zeta$ flavors are $\omega_\zeta$. }
\mylabel{fig:EffSchematic}
\end{figure}

An effective theory for the SD system with $\pi$-flux can be constructed in the limit of $\Omega \gg \Omega_c$ with the aid of \fig{fig:EffSchematic}. This figure brings out two important points: i) $\zeta = 1$ states make up the low energy manifold, and an effective theory can be constructed with the Hilbert space made only of the $\zeta=1$ states, ii) The special hopping structure is such that a $\zeta=1$ state at an odd site changes to a $\zeta = M$ state when it hops to a neighboring even site. As a result, to construct an effective low energy theory for the system in this limit, we can safely project out the $\zeta=M$ states to obtain a hopping only of the $\zeta = 1$ states. The $\zeta=1$ states at the odd sites then do not hybridize with those at the even sites -- the system has an emergent U($1$)$\times$U($1$) symmetry. Without loss of generality, we can then dub the odd site states as spin $\uparrow$ and even site states as spin $\downarrow$.

We define new fermionic operators for this effective ``spin-$\half$'' system by
\beq
b_{2i-1, 1} = a_{j, \uparrow} \,\,\,\, \text{and} \,\,\,\, b_{2i, 1} = a_{j, \downarrow} \,\,. 
\eeq
Here, $j$ labels the unit cells containing two neighboring sites labeled by $s$ ($\uparrow$ and $\downarrow$) as shown in \fig{fig:EffSchematic}. The onsite energy of the sites with $\zeta=1$ is $\omega_{1} = -2 \Omega \cos(\frac{\pi}{M+1}) \equiv -\epsilon$. The effective Hamiltonian of the system can then be written as
\beq
{\cal H}_{\mathrm{eff}} = H^{(0)} + H^{(1)} \,\,,
\eeq
where,
\beq
H^{(0)} = - \epsilon \sum_{j, s} n_{j, s} \,\,
\eeq
is the onsite energy term with the number operator defined as $n_{j, s} = a^{\dagger}_{j, s}  a_{j, s}$, and
\beq
H^{(1)} = - \sum_{j, s} \bigg[ \left(a^{\dagger}_{j+s, s} + a^{\dagger}_{j, s} \right) \frac{t^2}{(2\epsilon - U n_{j,\bar{s}})} a_{j, s} + \left(a^{\dagger}_{j, s} + a^{\dagger}_{j-s, s} \right) \frac{t^2}{(2\epsilon - U n_{j-s,\bar{s}})} a_{j, s} \bigg] \,\,.
\eeq
Here, we follow the notation of $s$ to be $+1$ ($-1$) for $\uparrow$ ($\downarrow$) particles and $\bar{s} = - s$.

Noting that, $n_{j,s} = 0$ or $1$ because of the fermionic nature of the particles, the following identity
\beq
\frac{1}{(2\epsilon - U n_{j,s})} = \frac{1}{2\epsilon} \left[ 1 + \frac{U n_{j,s}}{(2\epsilon - U)} \right] \,\,
\eeq 
holds. Using this identity, ${\cal H}_{\mathrm{eff}}$ can be recast into the form
\beq
\mylabel{eqn:Heffective}
{\cal H}_{\mathrm{eff}} = {\cal H}_0 + {\cal H}_1 + {\cal H}_2 + {\cal H}_3 \,\,,
\eeq
where,
\beq
{\cal H}_{0} = - \eff{\epsilon} \sum_{j, s} n_{j, s} \,\,,
\eeq 
\beq
{\cal H}_{1} = - \eff{t} \sum_{j, s} (a^{\dagger}_{j+s, s}  a_{j, s} + \mbox{h.~c.} ) \,\,,
\eeq
\beq
\mylabel{eqn:onsite}
{\cal H}_{2} = - \eff{U} \sum_{j, s} (n_{j, s} n_{j, \bar{s}} + n_{j+s, s}  n_{j, \bar{s}}) \,\,,
\eeq
\beq
\mylabel{eqn:current}
{\cal H}_{3} = - \eff{U} \sum_{j, s} (a^{\dagger}_{j+s, s}  a_{j, s} + \mbox{h.~c.} ) n_{j, \bar{s}} \,\,,
\eeq 
with, $\eff{\epsilon} = \left( \epsilon + \frac{t^2}{\epsilon} \right)$ is the effective onsite energy, $\eff{t} = \frac{t^2}{2\epsilon}$ is the effective hopping amplitude and $\eff{U} = \frac{t^2 U}{2 \epsilon (2 \epsilon - U)}$ is the effective interaction. 
We see that the interaction term (to leading order) has two physical contributions. The first \eqn{eqn:onsite} is a {\em non-local} interaction between particles at different physical sites of the optical lattice, and the second \eqn{eqn:current} is a ``correlated hopping'' term which is also found in other contexts~\cite{Liberto14,Cardarelli2016}.

\section{Field theory of the effective model at $\pi$-flux}
\mylabel{sec:Bosonization}
Closely following the refs.~\cite{Miranda03,Giamarchi04}, we now bosonize the effective Hamiltonian (\eqn{eqn:Heffective}) of the $\pi$-flux case in the large $\Omega$ limit constructed in the previous section. The bosonized Hamiltonian, containing only the forward scattering terms, can be written in the following form
\beq
\mylabel{eqn:Hboson}
{\cal H}^{0}_b = \sum_{\nu = \rho, \sigma} v_F (1 + {\bar{g}}_{4 \nu}) \sum_{q>0} q \bigg[ \sum_{\eta = R, L} {y^{\dagger}_{q,\eta,\nu}} y_{q,\eta,\nu} + x_\nu ({y^{\dagger}_{q,R,\nu}} {y^{\dagger}_{q,L,\nu}} + {\rm h.~c.}) \bigg] + {\rm const.} \,\,,
\eeq
where, $v_F = 2 t_{\mathrm{eff}} \sin(k_F)$ is the Fermi velocity and $k_F$ is the Fermi momentum of each of the species $s$. Here, $\eta$ labels the Right ($R$) and Left ($L$) movers, $\nu$ labels the two different sectors charge ($\rho$) and spin ($\sigma$). So, the spin-charge separation is clearly evident. The $y_{q,\eta,\nu}$-s are the bosonic annihilation operators, $x_\nu = \frac{\bar{g}_{2\nu}}{(1 + \bar{g}_{4\nu})}$ and $\bar{g}_{m'\nu} = \frac{g_{m'\nu}}{v_F}$ with $m' = 2$ and $4$ labeling the two types of forward scattering coupling constants. For this case, $g_{m' \rho} = - g_{m' \sigma} = \frac{f(k_F)}{\pi}$, where, $f(k_F) = V_0 + V_1 [1 + 2 \cos(k_F)]$ with $V_0 = V_1 = -U_{\mathrm{eff}}$. Using Bogoliubov transformation \cite{Miranda03}, the bosonized Hamiltonian in \eqn{eqn:Hboson} can be diagonalized and rewritten as 
\beq
\mylabel{eqn:Hbosonized}
{\cal H}^{0}_b = \sum_{\nu} \frac{u_\nu}{2} \int_{-\frac{L}{2}}^{\frac{L}{2}} dx \,\, \bigg[ K_\nu : \Pi^2_\nu(x) : + \frac{1}{K_\nu} : (\partial_x \phi_\nu(x))^2 : \bigg] \,\,,
\eeq
in terms of the dual bosonic field operators $\phi_\nu(x)$ and $\theta_\nu(x)$ with $\Pi_\nu(x) = \partial_x \theta_\nu(x)$. The pair $\phi_\nu$(x) and $\Pi_\nu(x)$ are canonically conjugate variables. Here, the Luttinger parameters ($K_\nu$) are given by
\beq
K_\nu = \sqrt{\frac{(1 + \bar{g}_{4\nu} - \bar{g}_{2\nu})}{(1 + \bar{g}_{4\nu} + \bar{g}_{2\nu})}} = \frac{1}{\sqrt{1-\nu \Lambda}} \,\,,\,\,\,\,\text{with}\,\,\,\,\,\,\Lambda = \frac{2 U \cot(\frac{k_F}{2})}{\pi (2\epsilon - U)} \,\,,
\eeq
following the notation of $\nu$ to be $+1$ ($-1$) for particles in the $\rho$ ($\sigma$) sector and their corresponding velocities are given by
\beq
u_\nu = v_F \sqrt{(1 + \bar{g}_{4\nu})^2 - (\bar{g}_{2\nu})^2} \,\,.
\eeq
We note that in this large $\Omega$ limit under consideration, the parameter $\Lambda > 0$ and hence $K_\rho > 1$ and $K_\sigma < 1$.

We now consider the effects of the back scattering terms. We are only interested in the physics of the system away from half filling hence the back scattering term is only operational in the spin sector. Then, the bosonized Hamiltonian containing this term can be written as (using \eqn{eqn:Hbosonized})
\beq
\mylabel{eqn:Hintbos}
{\cal H}_b = {\cal H}^{0}_b + \frac{2 g_1}{(2 \pi \alpha)^2} \int dx \,\, \cos(\sqrt{8\pi}\phi_{\sigma}(x))
\eeq 
with $\alpha$ being a microscopic length scale and $g_1$ being the coupling constant for the spin back scattering process. Here, we have,
\beq
g_1 = V_0 + V_1 [\cos(k_F) + \cos(2 k_F) + \cos(3k_F)] 
\eeq
and $g_1 \neq 0$ since $U_{\mathrm{eff}} \neq 0$.

Hence, by observing the facts that $K_\rho > 1$, $K_\sigma < 1$ and $g_1 \neq 0$, we reach the conclusions that charge is gapless and the spin is gapped (the spin back scattering term is relevant) \cite{Giamarchi04}. The system has dominant pair correlation $\Delta_s(r)$, defined in the text. It can also be written as
\beq
\mylabel{eqn:Delta}
\Delta_s(r) = \langle \hat{D}^{\dagger}_{j} \hat{D}_{j + r} \rangle, \,\,\,\, \hat{D}_{j} = a_{j, \uparrow} a_{j, \downarrow} \equiv b_{2i-1, 1} b_{2i, 1}
\eeq 
and thus corresponds to nonlocal correlation of squished pairs (nonlocal pairing in the optical lattice) $r$ distance away from each other. Closely following \cite{Miranda03}, we obtain
\beq
\mylabel{eqn:bosonizationDelta}
\Delta_s(r) \sim \frac{\beta_1}{r^{K_\rho}} - \frac{\beta_2 \cos(2 k_F r)}{r^{\left(K_\rho + \frac{1}{K_\rho} \right)}} \,\,,
\eeq 
where, $\beta_1$ and $\beta_2$ are positive parameters independent of $r$. This is a LE phase where low energy number changing excitations are $2$-particle type, and the central charge is $c=1$ as discussed in the text.

To compare these results with the numerical DMRG calculation, we calculate a pair correlation function, analogous to that in \eqn{eqn:Delta}, defined as
\beq
\mylabel{eqn:Delta_DMRG}
\Delta^{\text{DMRG}}_s(r) = \langle {\hat{O}_{i_0, s}}^{(2) \dagger} {\hat{O}_{i_0+r, s}}^{(2)} \rangle \equiv \langle b^{\dagger}_{i_0,1} b^{\dagger}_{i_0+1,1} b_{i_0+r+1, 1} b_{i_0+r, 1} \rangle \,\,,
\eeq 
where, $i_0 = L/2$ denotes the central site and the distance $r$ is even and takes values $0,2,4,\ldots,L/4$. In \fig{fig:bosonization} of the main text, we present the results of the comparison between the field theory and DMRG approaches for different values of $M$ and see excellent agreements. For the comparison, we have used $k_F=\frac{\pi}{2}Mn$ with $n=\frac{N}{LM}$ in the \eqn{eqn:bosonizationDelta} and the DMRG data of \eqn{eqn:Delta_DMRG} is fitted with the Bosonization result in \eqn{eqn:bosonizationDelta} using $\beta_1$ and $\beta_2$ as fitting parameters.

Finally, we point out the algebraic decay of $\Delta_s(r)$ more clearly by considering the $M=2$ case as an example and plotting the DMRG result of $\Delta_s(r)$ in the log-log scale shown in the \fig{fig:loglogpi}. Also, we define the following correlation function
\beq
\Delta_h (r) = \langle {\hat{O}_{i_0, h}}^{(2) \dagger} {\hat{O}_{i_0+r, h}}^{(2)} \rangle \equiv \langle b^{\dagger}_{i_0,1} b^{\dagger}_{i_0,2} b_{i_0+r, 1} b_{i_0+r, 2} \rangle \;.
\eeq
We now compare the DMRG results of $\Delta_s(r)$ and $\Delta_h(r)$ shown in \fig{fig:loglogpi} for $M=2$. We note that $\Delta_h(r)$ decays faster than $\Delta_s(r)$. Hence, the pair correlation function $\Delta_s(r)$ measuring the squishing effect is truly the dominant correlation of the SD system with $\pi$-flux.

\begin{figure*}[!ht]
{
\centering
\includegraphics[width=1.0\myfigwidth]{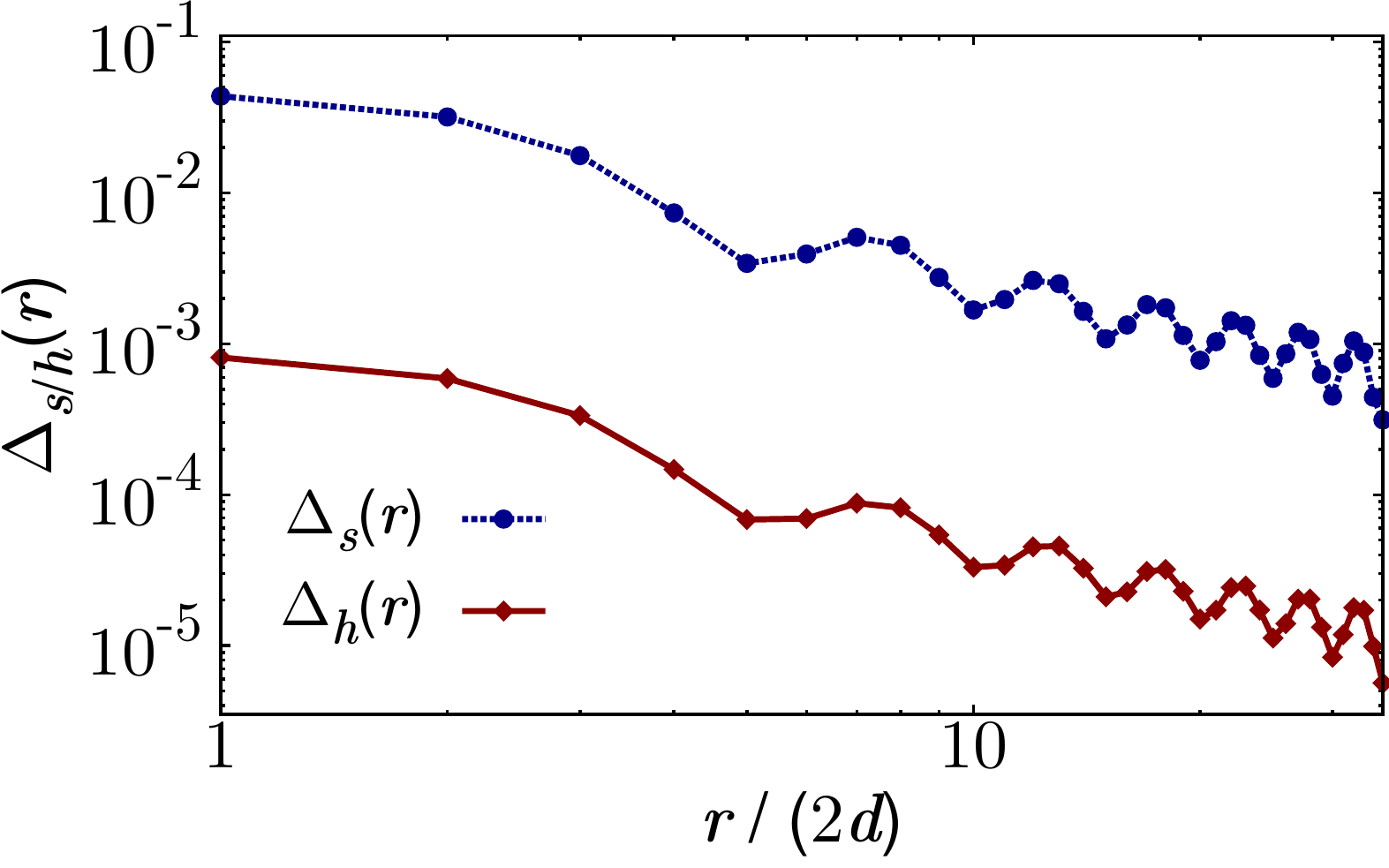}
}
\caption{(Color online) Comparison of the DMRG results of the two correlation functions $\Delta_s(r)$ and $\Delta_h(r)$ of the $M=2$ case for $\pi$-flux with $t/U=0.5$, $\Omega/U = 4$, $L=160$ and $N=32$.}
\mylabel{fig:loglogpi}
\end{figure*}

\begin{figure*}[!t]
{
\centering
\includegraphics[width=2.0\myfigwidth]{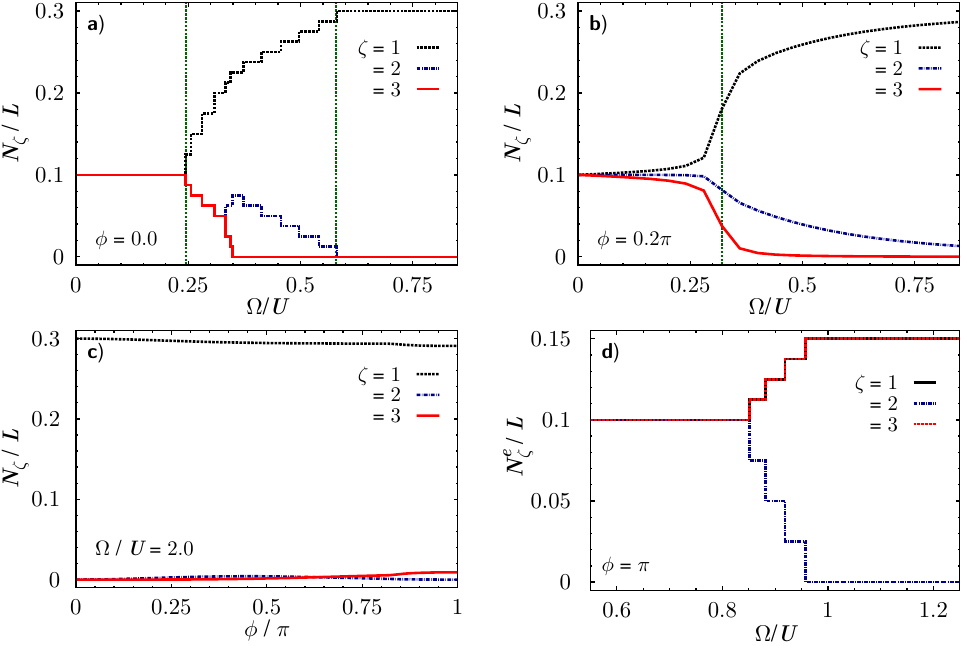}
}
\caption{(Color online) {\bf Population ($M=3$):} The variations of the total average populations ($N_\zeta$) of different $\zeta$-flavors are shown for the $M=3$ case with $t/U=0.5$, $n=0.1$ and $L=80$. Panel a) shows the variations of $N_\zeta$ as a function of $\Omega$ for $\phi=0$ case with the green lines showing the extent of the FFLO region. Similarly, in panel b) we show the same for the $\phi=0.2\pi$ case with the green line showing the transition point from the BF state to the FM state. While in panel c) we show the variations of $N_\zeta$ as a function of $\phi$ in the $\Omega \gg \Omega_c$ with $\Omega/U=2$. Finally, for the special case $\phi=\pi$, we show the variations of the effective populations $N^e_\zeta$ as a function of $\Omega$ in the panel d).}
\mylabel{fig:popu}
\end{figure*}

\section{Population of different flavors}
\mylabel{sec:population}

The nature of different phases of the SD system discussed in the main text can be further investigated by looking at the populations of different $\zeta$ flavors. The total average population $N_\zeta$ of a particular $\zeta$ flavor is defined as 
\beq
N_\zeta = \sum_i \langle n_{i,\zeta} \rangle \;,
\eeq
with $\langle n_{i,\zeta}\rangle$ being the average occupation of the $\zeta$ flavor at the site $i$. The special cases $\phi = 0$ and $\pi$ have $M$ number  of $U(1)$ symmetries corresponding to particle number conservations of each of the $M$ individual $\zeta$ flavors. It is noteworthy that the $\phi=0$ case with $M=2$ is equivalent to the attractive Hubbard model in the presence of a Zeeman field and this model has an FFLO region~\cite{Feiguin2011} corresponding to ``intermediate'' strengths of the Zeeman field. 

We show the variations of $N_\zeta$ in \fig{fig:popu}(\textbf{a}--\textbf{c}) for the $M=3$ case for three different values of $\phi$. In \fig{fig:popu}\textbf{a}, we note that at $\phi=0$ for small $\Omega$, the populations of all the $\zeta$ flavors are equal indicating the presence of the $3$-BF state. With increasing $\Omega$, there is a regime of $\Omega$ over which populations of different flavors are unequal but non-zero, i.~e. there is imbalance between different flavors. This regime corresponds to the FFLO state. With further increase in $\Omega$, the state becomes an FM state when all the the states are of $\zeta = 1$ flavor. Corresponding excitation spectra of this case is shown in \fig{fig:M3}\textbf{d} of the main text. The plateaus in the FFLO regime are due to the $3$ $U(1)$ symmetries special to this case and finite system size. In the contrary, for the $\phi=0.2\pi$ case shown in \fig{fig:popu}\textbf{b}, there is a direct transition from the $3$-BF state to the FM state without any intervening FFLO phase as also seen in the phase diagrams \fig{fig:M3}(\textbf{a}--\textbf{c}) shown in the main text. Then, we concentrate on the interesting limit of $\Omega \gg \Omega_c$ and show the variations of $N_\zeta$ as a function of $\phi$ for $\Omega/U = 2$ in \fig{fig:popu}\textbf{c} (corresponding excitation spectra is shown in \fig{fig:M3}\textbf{e} of the main text). We note that in this limit the population of the $\zeta = 1$ flavor is always the largest whereas $N_2 \approx N_3 \approx 0$.

The idea of constructing an effective theory for the $\pi$-flux case in the $\Omega \gtrsim \Omega_c$ limit (discussed in the previous two sections \sect{sec:EffTheory} and \sect{sec:Bosonization}) becomes clearer by looking at the variations of the effective populations $N^e_\zeta$ of different $\zeta$ flavors. As an example, for the $M=3$ case they are defined as,
\bea
N^e_1 &=& \sum_{i=1}^{L/2} (\langle n_{2i-1,1} \rangle + \langle n_{2i,3} \rangle) \,\,,\\
N^e_2 &=& N_2 \,\,,\\
N^e_3 &=& \sum_{i=1}^{L/2} (\langle n_{2i-1,3} \rangle + \langle n_{2i,1} \rangle) \,\,.
\eea
Variations of $N^e_\zeta$ as a function of $\Omega$ is shown in \fig{fig:popu}\textbf{d} for the $M=3$ case with $\pi$-flux. Interestingly, at small $\Omega$, all the effective populations are equal but at large $\Omega$, $N^e_1 = N^e_3$ and $N^e_2 = 0$. These two states correspond to the BF state and the SBF state respectively. The U($1$)$\times$U($1$) symmetry is thus evident in the SBF state. We also note that for intermediate $\Omega$, there is a regime over which $N^e_2 \neq 0$ but $N^e_2 \neq N^e_1 = N^e_3$. In this regime, the $3$-baryons are destabilized in favor of the $2$-baryons. Again, the plateaus in the intermediate regime are due to the $3$ $U(1)$ symmetries special to this case and finite system size.

\section{Lifshitz Transitions}
\mylabel{sec:lifshitz}

\begin{figure*}[!ht]
{
\centering
\includegraphics[width=1.0\myfigwidth]{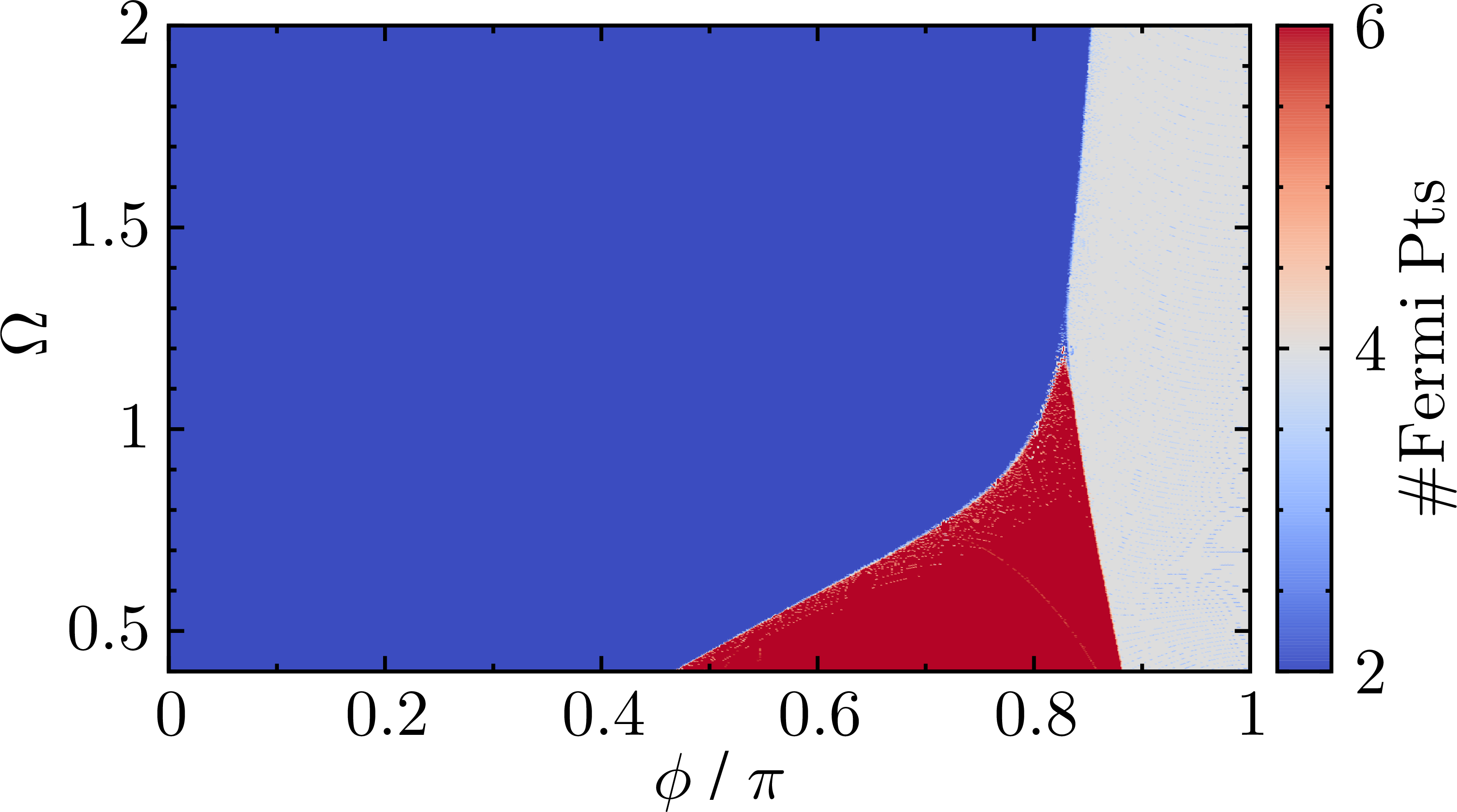}
}
\caption{(Color online) Lifshitz transitions occurring in the $\Omega$-$\phi$ plane for the non-interacting $M=3$ SD system with $t=0.5$ and $n=0.1$ are shown here by looking at the number of Fermi points in the Fermi surface for the given density.}
\mylabel{fig:Lifshitz}
\end{figure*}

The topology of the Fermi surface plays an important role in understanding manybody phases obtained here. A change in the topology of the Fermi surface, so called Lifshitz transition~\cite{Lifshitz30}, has spectacular consequences in the physical properties of fermionic systems such as the resistivity and thermo-electric power~\cite{Blanter94,Silaev15}, superconductivity~\cite{Norman10,Leonov15}, specific heat and lattice dynamics~\cite{Potzel95,Katsnelson94,Novikov99} etc. Here, we show that several Lifshitz transitions (LT) take place in the non-interacting SD system in the $\Omega$-$\phi$ plane. As expected, dressed by the interaction they produce novel consequences in the manybody phase diagrams (see \fig{fig:M2} and \fig{fig:M3} of the main text) of the SD system.

The LT from a Fermi surface having $2$ Fermi points to that having $4$ Fermi points for the $M=2$ SD system gives the transition line corresponding to the transition from an FM state to an SBF state as discussed in \fig{fig:M2}\textbf{h} of the main text. As also discussed in the main text and shown in \fig{fig:M3} there, the situation is much more interesting for the $M=3$ case. For this case, we note that there are several LTs occurring as a function of $\phi$ and $\Omega$ as shown in the \fig{fig:Lifshitz}. Similar to the $M=2$ case, the transition from the FM state in the large $\Omega$ limit is associated with the LT of the Fermi surface having $2$ Fermi points to that having $4$ Fermi points. There are also interesting LTs from $2$ to $6$ to $4$ Fermi points but the precise characterization transitions of different phase using this LTs for this case is complicated by the intervening extended phase separation region.

\section{Bound states in different limiting cases}
\mylabel{sec:2particle}

In this section, we discuss the properties of different bound states formed in the SD system in the following two interesting limits.

$1$) $\frac{\Omega}{U} \gg \frac{t}{U}$ and general $\phi$: In order to further analyze the interplay between the Lifshitz transitions and the emergence of the SBF phase, we now analyze the effect of the interaction on the level of two-particle physics, which has been shown to be a very useful approach for describing various systems such as bosons in the dilute limit~\cite{Mishra2015}, or spin systems close to the saturation magnetic field~\cite{Kecke2007,Kolezhuk2012,Azimi2014}. We can get a quasi-analytical picture of the two-particle physics of the system in this limit by constructing an effective theory in the low energy sector consisting only of the $\zeta=1$ states. Up to second order perturbation theory, we obtain an effective spinless fermion model only in the $\zeta=1$ manifold with the nearest and next nearest neighbor tunneling amplitudes and interaction terms:
\begin{align}
\mathcal{H}_{\mathrm{eff}}^\phi= t_1 \sum_i \left( b_{i,1}^\dagger b_{i+1,1} +{\rm H.c.}\right) 
+ t_2 \sum_i \left( b_{i,1}^\dagger b_{i+2,1} +{\rm H.c.}\right) 
+ t_2^c \sum_i \left(b_{i,1}^\dagger n_{i+1,1} b_{i+2,1}+{\rm H.c.}  \right)
+ V \sum_i n_{i,1} n_{i+1,1} \;,
\end{align}
where,
\begin{align}
t_1=\dA_{1,1} \;,\quad
t_2=\sum_{\zeta=2}^{M} \frac{\dA_{1,\zeta}^2}{ \domg_1 - \domg_\zeta} \;,\quad
t_2^c=-\sum_{\zeta=2}^{M} \frac{\dA_{1,\zeta}^2 U }{ (\domg_1 - \domg_\zeta) (\domg_1 - \domg_\zeta - U)} \;,\quad
V=\sum_{\zeta=2}^{M} \frac{|\dA_{1,\zeta}|^2 U }{ (\domg_1 - \domg_\zeta) (\domg_1 - \domg_\zeta - U)} \;.
\end{align}

Starting from a general two-particle state which is given by $\left| \Psi \right> = \sum_{i,j>i} c_{i,j} b_{i,1}^\dagger b_{j,1}^\dagger \left|0\right>$, the Schr\"odinger equation can be written as $H_{\mathrm{eff}}^\phi \left|\Psi\right> = E \left|\Psi\right>$, where $E$ is the two particle energy. Expressing the amplitudes as $c_{i,i+r} = c_r {\rm e}^{i Q (i+\frac{r}{2})}$  
leads to the following system of coupled equations for $c_r$:
\begin{eqnarray}
&&\!\!\!\!\!\!\!\!\!\!(E-V)c_1 = -2  t_1 \cos\left(\frac{Q}{2}\right) c_2 + 2 t_2^c \cos(Q) c_1 - 2 t_2 \cos(Q) c_3 \;, \mylabel{eqn:dilute_2particle_eqs-2} \\
&&\!\!\!\!\!\!\!\!\!\! E c_2 = -2  t_1 \cos\left(\frac{Q}{2}\right) \left( c_1 + c_3 \right) - 2 t_2 \cos(Q) c_4 \;, \mylabel{eqn:dilute_2particle_eqs-3}\\
&&\!\!\!\!\!\!\!\!\!\! E c_r = -2  t_1 \cos\left(\frac{Q}{2}\right) \left( c_{r-1} + c_{r+1}\right) - 2 t_2 \cos(Q) \left( c_{r-2} + c_{r+2}\right)\; , \; r\geq 3 \;. \mylabel{eqn:dilute_2particle_eqs-4}
\end{eqnarray}
For two scattering particles with momenta $k_1$ and $k_2$, $Q=(k_1+k_2)$ is the total center of mass momentum and $q=\frac{k_1-k_2}{2}$ is the relative momentum. The energies of the two body bound states are determined by the interaction and correlated hopping terms. Generally, we look for bound state solutions with the ansatz $c_{r}=\alpha^r \;, r\geq 3$
with $|\alpha|<1$. Solving the set of \eqn{eqn:dilute_2particle_eqs-2}-\eqn{eqn:dilute_2particle_eqs-4}, we obtain $\alpha$, $c_1$ and $c_2$.

\begin{figure*}[!ht]
{
\centering
\includegraphics[width=2.2\myfigwidth]{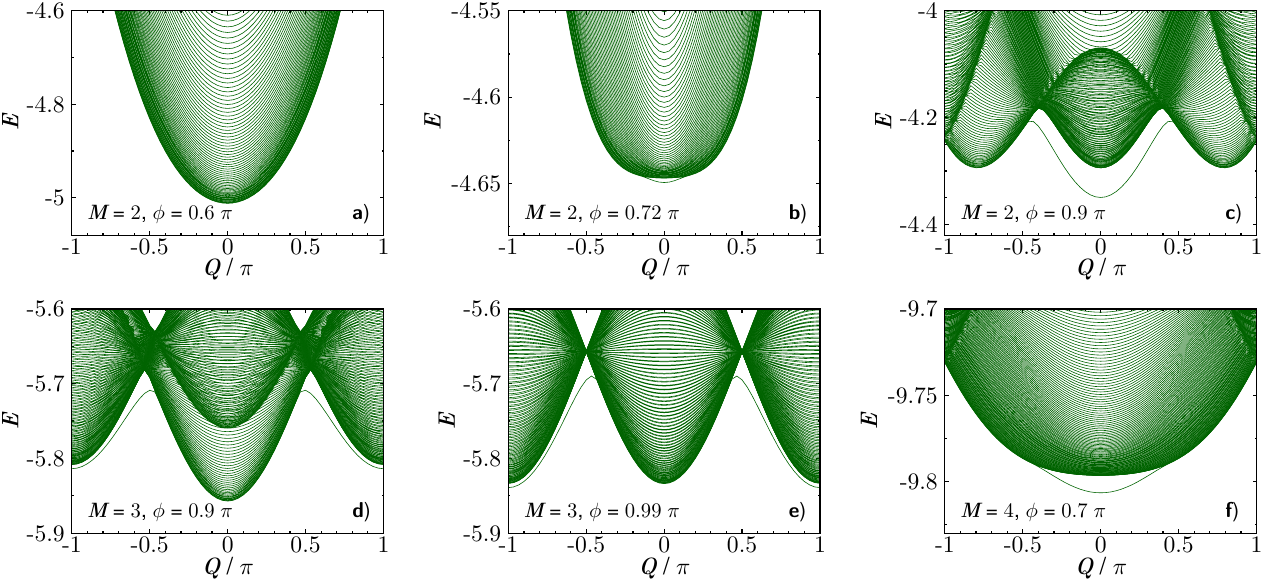}
}
\caption{Low energy part of the two-particle scattering spectrum $E(Q)$ as obtained from the solutions of \eqn{eqn:dilute_2particle_eqs-2}-\eqn{eqn:dilute_2particle_eqs-4}. For $M=2$ and $\Omega/U=2$ ($\phi_\mathrm{LT} \approx 0.76\pi$), panels a), b) and c) show the cases $\phi=0.6\pi$, $0.72\pi$ and $0.9\pi$ respectively. For $M=3$ and $\Omega/U=2$ ($\phi_\mathrm{LT} \approx 0.85\pi$), panels d) and e) show the cases $\phi=0.9\pi$ and $0.99\pi$ respectively. The last panel f) shows the case $\phi=0.7\pi$ for $M=4$ and $\Omega/U=3$ ($\phi_\mathrm{LT} \approx 0.72\pi$).}
\mylabel{fig:2scattering}
\end{figure*}

In \fig{fig:2scattering}, we present the low energy part of the two-particle spectrum as solutions of \eqn{eqn:dilute_2particle_eqs-2}-\eqn{eqn:dilute_2particle_eqs-4}. The isolated solid line indicates the presence of a bound state. \Fig{fig:2scattering}(\textbf{a}-\textbf{c}) indicate that for the $M=2$ case, as stated in the text, very close to the Lifshitz transition at $\phi=\phi_\mathrm{LT}$ in the large $\Omega$ limit, where the scattering part of the spectrum acquires a second minimum, also a bound state emerges at $Q=0$ (\fig{fig:2scattering}\textbf{b} and \textbf{c}). From the fact that this bound state immediately becomes of the lowest energy, we can understand why for small filling we observe a direct transition from the FM phase to the SBF phase. 

This situation is different for the case of $M=3$ (\fig{fig:2scattering}\textbf{d} and \textbf{e}) where the bound state becomes of the lowest energy only for ``large enough'' flux $>\phi_\mathrm{LT}$. Hence, we may expect an intermediate multicomponent partially paired phase between FM and SBF phases as observed in our numerical results discussed in the phase diagrams \fig{fig:M3}(\textbf{a}--\textbf{c}) of the main text.

For $M=4$, however, we again observe the emergence of the lowest-energy bound state for smaller flux $<\phi_\mathrm{LT}$ (\fig{fig:2scattering}\textbf{f}) and, hence, a direct transition between FM and SBF phases without an intermediate phase may be expected for small fillings. This interesting phenomenon has to be examined by detailed numerical simulations.

$2$) Dilute limit ($n \ll 1/M$): In order to obtain a full picture in the dilute limit ($n \ll 1/M$), we evaluate the energies of multi-particle bound states of the SD system. We numerically calculate the energies of the two-, three- and four-particle states in an empty lattice by means of DMRG simulations for $M=2, 3$ and $4$ respectively which by comparison to the single particle energies allow us to estimate positions of different transitions. In \fig{fig:dilute_bound}, we depict the transition lines as well as the large $\Omega$ part of the Lifshitz transition line. These estimates of the transitions are shown as different lines in figs.~(3) and ~(4) of the text. They compare well to the numerical estimates from excitation spectra shown in \fig{fig:M2} and \fig{fig:M3} of the main text in the low filling ($n=0.1$) case considered (in particular for $\phi \approx 0$ and $\phi \approx \pi$). 

\begin{figure*}[!ht]
{
\centering
\includegraphics[width=2.2\myfigwidth]{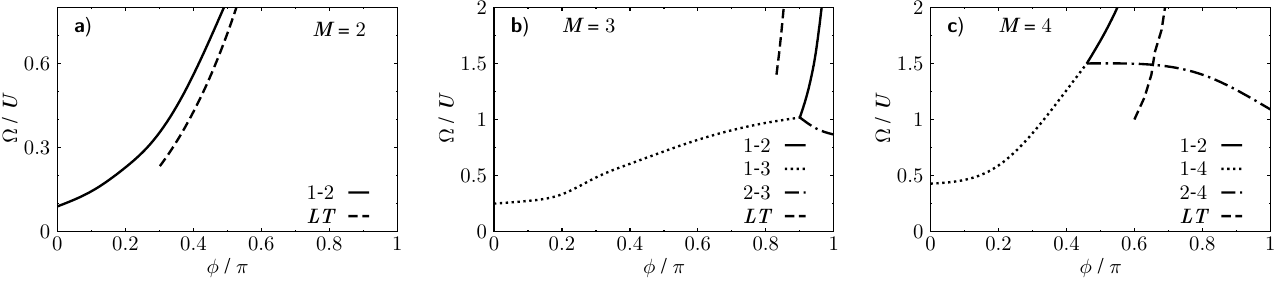}
}
\caption{(Color online) Phase diagram in the dilute limit $n \approx 0$ for a) $M=2$, b) $M=3$ and c) $M=4$. Solid, dotted and dash-dotted lines indicate various crossings between different few-particle bound state energies to be the lowest energy for the few-particle system. For the sake of completeness, we add the Lifshitz-transition lines (dashed lines) in the large $\Omega$ limit.}
\mylabel{fig:dilute_bound}
\end{figure*}

\section{Numerical analysis}
\mylabel{sec:numerical}

In this section, we briefly discuss our numerical analysis using the DMRG of characterizing different phases and their transitions. We focus on the regime of low density ($n\leq 1/M$). The bound state phases and the FM phase are best distinguished by considering the lowest energy particle number changing $m$-particle excitations ($\Delta E_m$) in the canonical ensemble defined by
\beq
\mylabel{eqn:excispec}
\Delta E_m = \frac{E_0(N+m,L)+E_0(N-m,L)-2 E_0(N,L)}{2 m} \;,
\eeq
where, $E_0(N,L)$ is the ground state energy of an SD system with $L$ optical lattice sites and $N$ particles. While in the FM phase the single particle excitation $\Delta E_1$ is the lowest; the bound state phases, BF and SBF, are characterized by $\Delta E_2$ and $\Delta E_3$ respectively becoming the lowest gap in the finite size system. The crossings between these excitations mark different transition points. We may also relate the transitions from bound to the FFLO phases by $\Delta E_2=\Delta E_1$ or $\Delta E_3=\Delta E_1$.

\begin{figure}[!ht]
{
\centering
{\includegraphics[width=1.0\myfigwidth]{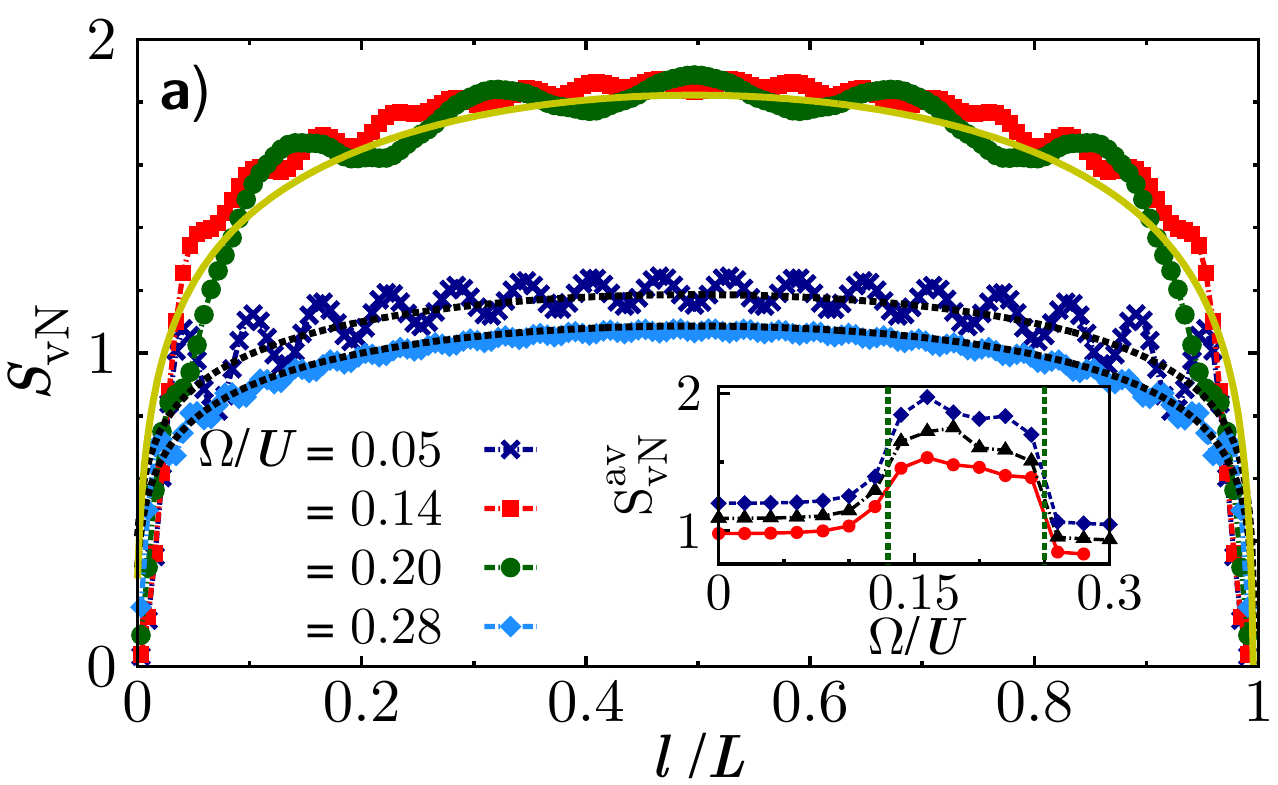}~~~~~~~~~\includegraphics[width=1.0\myfigwidth]{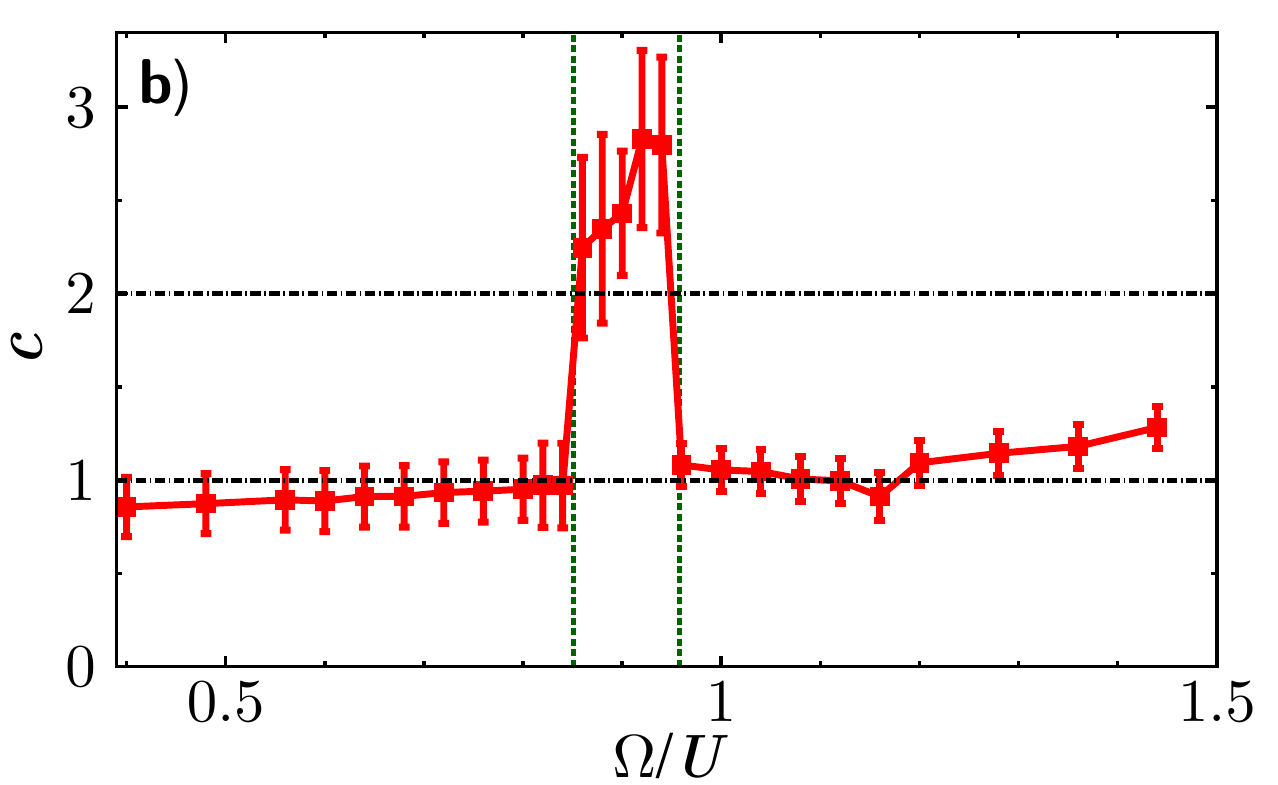}}
}
\caption{(Color online) a) Entanglement entropy $S_{\mathrm{vN}}$ as a function of the subsystem size $l$ for the $M=2$ case with $\phi=0.05\pi$, $t/U 0.5$ and several values of $\Omega$. The black dotted lines are fits to the Calabrese-Cardy-formula (\eqn{eqn:CC}) with $c=1$ in the BF ($\Omega/U=0.05$) and FM ($\Omega/U=0.28$) phases depending on the data, while the solid yellow line is the same with $c=2$ in the FFLO region ($\Omega/U=0.14$ and $\Omega/U=0.2$). The inset shows the average entanglement entropy $\mathrm{S}_{\mathrm{vN}}^{\mathrm{av}}$ as a function of $\Omega/U$ for system sizes $L=40$, $80$ and $160$ from bottom to top and the green dotted lines denote the extent of the FFLO region as determined by the excitation spectra defined in \eqn{eqn:excispec}. b) The estimated central charge ($c$) using similar procedure as a) for the $M=3$ case with $\phi=\pi$ and $t/U 0.5$ as a function of $\Omega/U$ ($L=80$). Corresponding excitation spectra is shown in \fig{fig:M3}\textbf{f} of the main text.  While for the BF ($\Omega/U\lesssim 0.85$) and the SBF ($\Omega/U\gtrsim 0.95$) phases we observe $c=1$, the fitting result for the intermediate imbalanced phase may be consistent with $c=2$. The green dotted lines again denote the extent of the imbalanced region as determined by the excitation spectra.  
}
\mylabel{fig:numerics_ee_sv}
\end{figure}

The most important tool, however, to characterize various gapless phases with different central charge ($c$) is the von-Neumann block-entanglement entropy $S_{\mathrm{vN}}= -\mathrm{tr}\left( \rho_l \ln\rho_l \right)$, where $\rho_l$ is the reduced density matrix of a subsystem of length $l$ embedded in a chain of a 
finite length $L$. For conformally invariant gapless states~\cite{Vidal2003, Calabrese2004}, one can relate the scaling of the entanglement entropy to the central charge of the system.
In particular, for different subsystem lengths of $1$D systems with open boundaries, $S_{\mathrm{vN}}(l)$ is described by the Calabrese-Cardy (CC) formula \cite{Calabrese09}
\beq
S_{\mathrm{vN}}(l) = \frac{c}{6} \ln\left[ \frac{L}{\pi} \sin\left(\frac{\pi}{L}l\right) \right] + \cdots \,\,,
\mylabel{eqn:CC}
\eeq
The ellipsis contains non-universal constants and higher order oscillatory terms due to the finite system size (see \fig{fig:M3}\textbf{g} of the main text and \fig{fig:numerics_ee_sv} for examples). The entanglement entropy and also the spectrum of eigenvalues of the reduced density matrix themselves have been shown to offer a sensitive probe for quantum phase transitions~\cite{Calabrese2004, Li08, Pollmann10, Poilblanc10}. We also compute the averaged entanglement entropy $\mathrm{S}_{\mathrm{vN}}^{\mathrm{av}}$
 over all bi-partitions ($l$) in the range $[L/4, 3L/4]$ reducing the finite size effects and it increases in the FFLO regime as shown in the inset of \fig{fig:numerics_ee_sv}\textbf{a}. The central charge ($c$) is estimated directly by fitting the numerical results with the \eqn{eqn:CC} as shown for example in \fig{fig:numerics_ee_sv}\textbf{b}. However, for finite systems with open boundary conditions and in particular for low fillings, strong oscillatory terms complicate this analysis. Hence, the central charge in the intermediate imbalanced phase is overestimated due to the finite system size effects (compare \fig{fig:numerics_ee_sv}\textbf{b}).

\begin{figure*}[!ht]
{
\centering
\includegraphics[width=2.0\myfigwidth]{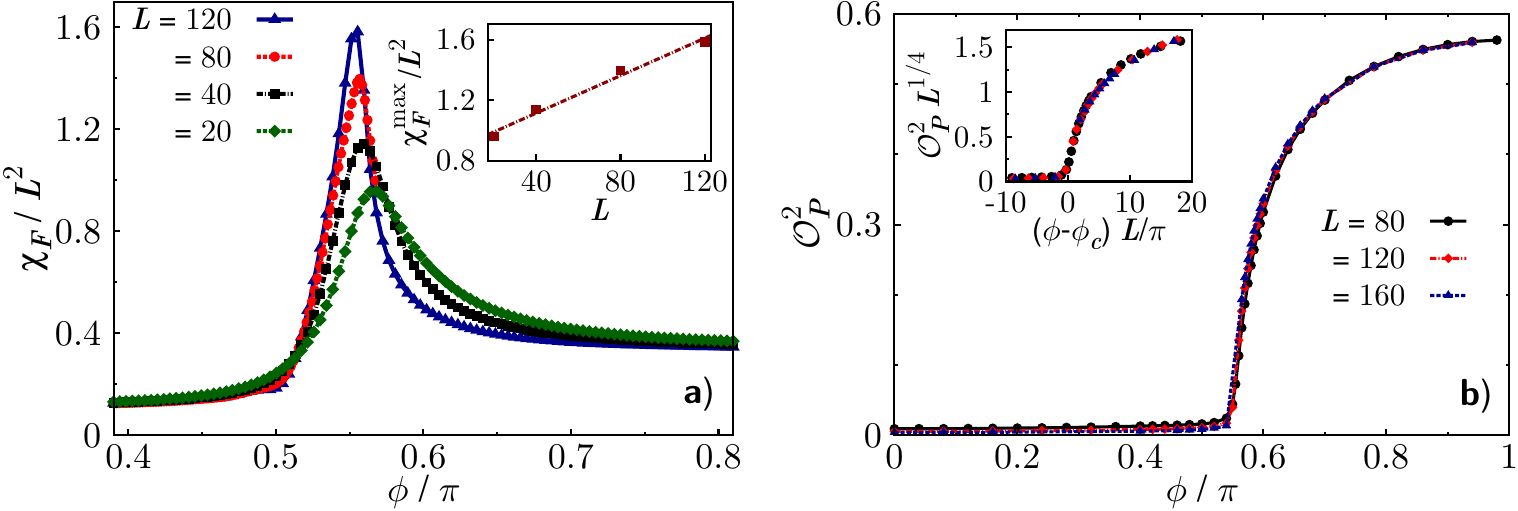}
}
\caption{(Color online) Analysis of the transition from FM to SBF phase as a function of $\phi$ for $M=2$, $\Omega/U=0.8$, $n=0.1$. (a) Behavior of the fidelity susceptibility $\chi_F/L^2$ as a function of $\phi$. The inset shows the linear scaling of the peak height ($\chi^{max}_F/L^2$) with the system size. (b) Parity-order parameter $\mathcal{O}_P^2$ as a function of $\phi$ is shown. The collapse of the data points is shown in the inset ($\phi_c \approx 0.55\pi$).}
\mylabel{fig:IsingM2}
\end{figure*}

Finally, we present details of the transition between the FM and SBF phases for the $M=2$ case in the large $\Omega$ limit as a function of $\phi$. Contrary to the $M=3$ case, we observe a single critical point between the two phases without crossing to an intermediate phase. In \fig{fig:IsingM2}, we present data for a cut same as the \fig{fig:M2}\textbf{g} (showing the corresponding excitation spectra) of the main text for $\Omega/U=0.8$, which gives strong indications of the presence of a ``critical'' point. Indeed, similar critical points between Luttinger liquids of bound and unbound pairs have been discussed, for example in spin-systems~\cite{Chen2003} or bosons~\cite{Daley2009,Greschner2013}.

In \fig{fig:IsingM2}\textbf{a}, we show the scaling of the fidelity susceptibility~\cite{gu2010}
\beq
\chi_{F}(\phi) = \lim_{\delta\phi\to 0} \frac{-2 \ln |\langle \Psi_0(\phi) | \Psi_0(\phi + \delta\phi) \rangle| }{(\delta\phi)^2}
\eeq
with $|\Psi_0\rangle$ being the groundstate wavefunction. As the system size increases, $\chi_{F}/L^2$ develops a distinct divergent single peak and the peak-height ($\chi^{max}_F/L^2$) scales linearly with respect to its wings~\cite{Greschner2013b,Azimi2014}.
Note that since we are calculating the fidelity-susceptibility with respect to the flux one generally observes a quadratic scaling of the fidelity susceptibility with the system size $\chi_{F}\sim L^2$~\cite{Greschner2013b,Azimi2014}. Equivalently, for a crossing of the transition at fixed flux (data not shown), we observe that $\chi_{F}\sim L$ within the FM and SBF phases and $\chi_{F}\sim L^2$ for the transition.

An ``order parameter'' of this transition can be constructed and it is given by the parity order
\beq
\mathcal{O}_P^2 (i,j) = {\rm e}^{i \sum_{i<k<j} \sum_\zeta \pi n_{k,\zeta}} \;.
\eeq
In \fig{fig:IsingM2}\textbf{b}, we plot the averaged parity order $\mathcal{O}_P^2 = \frac{2}{L} \sum_{L/4<j<3L/4} \langle \mathcal{O}_P^2 (L/4,j)\rangle$ as a function of the flux for several system sizes. The parity order vanishes in the FM phase $\mathcal{O}_P^2\to 0$ while it becomes finite in the SBF phase and there is an interesting scaling as illustrated in the inset of \fig{fig:IsingM2}\textbf{b}.

\section{Phase Separation}
\mylabel{sec:PS}

\begin{figure*}[!ht]
{
\centering
\includegraphics[width=2.0\myfigwidth]{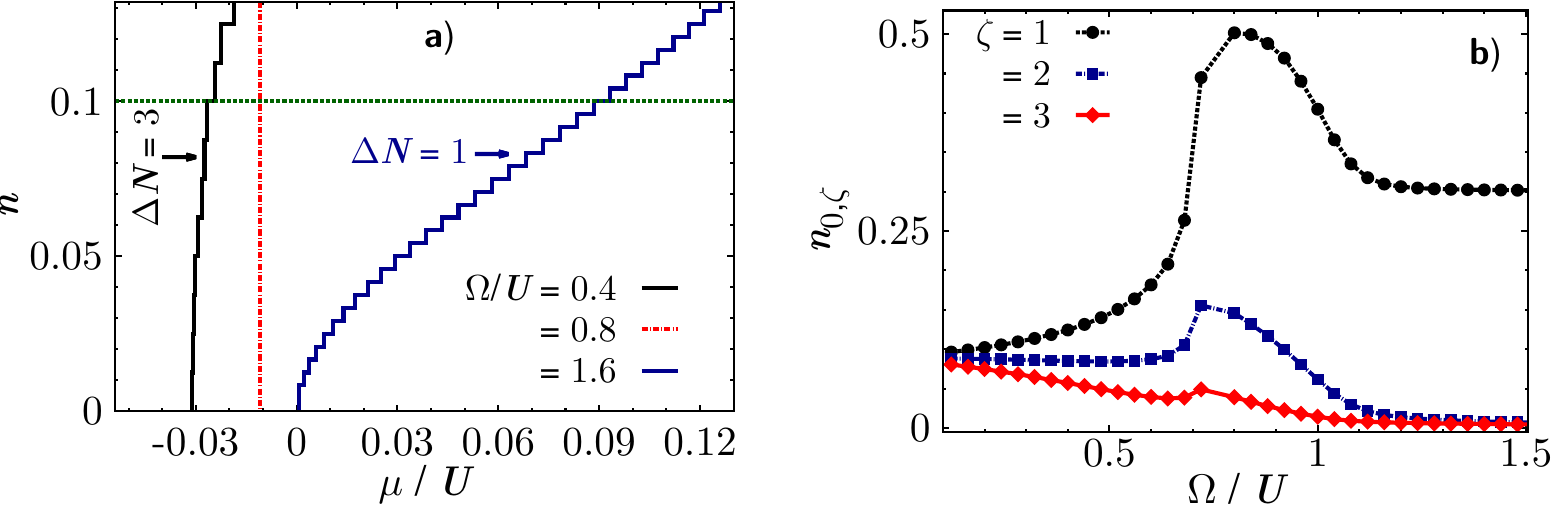}
}
\caption{(Color online) {\bf Phase separation ($M=3$):} Considering a particular flux $\phi/\pi = 0.6$, we show phase separation in the $M=3$ SD system for $t/U=0.5$, $n=0.1$ and $L=80$. In a) we show the behavior of the equation of state of the system, i.~e. the density ($n$) as a function of the chemical potential ($\mu$), for different values of $\Omega$. The chemical potential $\mu$ has been shifted for clarity. The green line shows the density $n=0.1$ under consideration and the $\Delta N$-s show the steps in the number of particle $N$. In b) we show the variation of the average occupation at the central site $n_{0,\zeta}$ as a function of $\Omega$.}
\mylabel{fig:PS}
\end{figure*}

The propensity of phase separation (PS) for the $M=3$ case is further analyzed in this section. 
In \fig{fig:PS}\textbf{a}, we show the equation of state $n(\mu)$, where $\mu$ is the chemical potential of the system as obtained from DMRG calculations by the minimization of the functional $E_0(N,L)-\mu N$~\cite{Mishra2015}. While the BF and FM phases are characterized by a series of steps in the number of particles $N$, $\Delta N=3$ and $\Delta N=1$ respectively~\cite{Azimi2014}, we note that there exists a regime where $\mu=$ constant, i.~e. a small change in the chemical potential produces a large change in the density. Hence, according to the definition of the compressibility $\kappa^{-1} = n^2 \left( \frac{\partial \mu}{\partial n}\right)$, the system is infinitely compressible in this regime. As a result, as discussed in the text, particles tend to cluster near the central site at $L/2$ and there is PS in the system. We further note from the behavior of the average occupation at the central site $n_{0,\zeta} \equiv \langle n_{i_0,\zeta} \rangle$ shown in \fig{fig:PS}\textbf{b} that with the increase in $\Omega$ from zero, there is a dramatic increase in $n_{0,\zeta}$ for $\Omega$ values in the PS regime.

The phenomenon of PS can also be understood by looking at the non-interacting single particle band structure of the system. It is noted that in the PS region, the energy bands of the system become extremely flat (for a discussion of interesting phenomena occurring in systems with flat bands see also the refs.~\cite{Huber2010,Tovmasyan2016}). As a result, the kinetic energy of the system is frustrated and the system can only gain energy from the non-local induced interaction energy (in the similar way discussed in \sect{sec:EffTheory}) by clustering the particles. The system thus phase separates with high density of particles near the central site. We expect PS to occur for $M > 3$ SD systems as well.

\section{Chiral Currents}
\mylabel{sec:chiral}
Recent experimental works on ladder systems~\cite{Atala2014} and SD systems~\cite{Stuhl2015,Mancini15} focused on the study of chiral (also called boundary or edge) currents. Also theoretically the chiral currents have been suggested to be very interesting observables which characterize various quantum phases~\cite{Piraud2015, Greschner2015, Kolley2015, Barbarino2016, Cornfeld2016}. From this viewpoint, we now discuss the behaviors of the chiral currents for the SD system with \SU{M} symmetric attractive interactions.

\begin{figure}[!ht]
\centering
{
\centering
\includegraphics[width=2.2\myfigwidth]{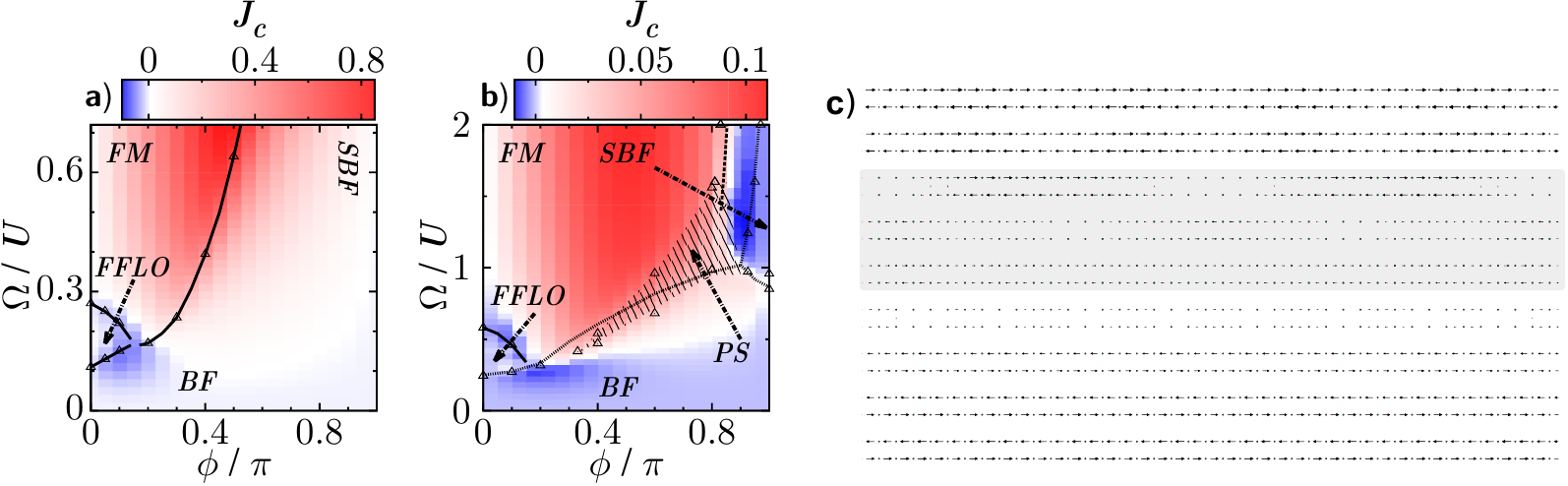}
}
\caption{(Color online) Chiral current $J_c$ with $n=0.1$ and $t/U=0.5$ for a) $M=2$ ($L=80$ rungs) and b) $M=3$ ($L=40$ rungs). Solid lines are guides to the eye, dotted lines are the numerical estimates of bound state transitions in the dilute limit and the dashed lines correspond to a LT. c) Examples of the local current configurations in the original basis (Eqn.~($1$) of the main text) of $M=2$ with $\phi/\pi=0.05$ and (top to bottom) $\Omega / U =0.08$, $0.12$, $0.16$, $0.2$, $0.24$, $0.28$, $0.36$, $0.44$ and $0.5$. The shading marks the current configurations corresponding to the c=2 phase.
}
\mylabel{fig:pd_jc}
\end{figure}

From the continuity equation, we can define a current $J({\bf r}\to {\bf s})$ from a site ${\mathbf r}$ to a neighboring site ${\mathbf s}$ for a system with the Hamiltonian ${\cal H}$ as 
$\left<\frac{dn_{{\bf r}}}{dt}\right>= {\rm i}\left<[\mathcal{H}, n_{\bf r}]\right> =-\sum_{\left<{\bf s}\right>} J({\bf r}\to {\bf s})$,
where $n_{{\bf r}}$ is the density at site {\bf r}. In particular, in the \textit{original basis} (see Eqn.~($1$) of the main text) we define current for the SD system as
\begin{align}
J\left((i,\gamma) \to (i+1,\gamma)\right) &= {\rm i} t \left( C^\dagger_{{i+1,\gamma}} C_{i,\gamma} - C^\dagger_{i,\gamma} C_{{i+1,\gamma}} \right) \;.
\end{align}
The configuration of the local currents can give important insights into the properties of a quantum phase, but the average current that circulates through the boundary of a system, called chiral current, can give more interesting information. From the spin-resolved momentum distribution, this chiral current can be measured experimentally in typical setups~\cite{Stuhl2015, Mancini15} and is defined as
\beq
J_c =\frac{1}{L}\sum_{i} \left[\langle J\left((i,1) \to (i+1,1)\right) \rangle - \langle J\left((i,M) \to (i+1,M)\right) \rangle \right]\;.
\eeq
Using the Hellman-Feynman theorem, we can compute $J_c$ from a derivative of the ground state energy~\cite{Piraud2015, Greschner2015} $E_0$ and for the SD system
\beq
J_c^{M=2}=\partial_\phi E_0 / L \quad \text{ and } \quad J_c^{M=3}=2\partial_\phi E_0 / L \;.
\eeq

In \fig{fig:pd_jc}(\textbf{a} and \textbf{b}), we present the chiral currents for $M=2$ and $M=3$ respectively, corresponding to fig.~($3$) and ($4$) of the text. For both the cases, $J_c$ is strongly suppressed in the bound state phases. In the FM region, it increases approximately linearly with the flux and stays roughly constant as a function of $\Omega/U$. This feature allows us to interpret the FM-phase to be a Meissner-like phase~\cite{Orignac2001}. We may identify some of the phase transitions by the kinks in the chiral current. In particular, during the FM to SBF transition, $J_c$ exhibits a strong drop and becomes slightly negative in the the SBF-phase.

\Fig{fig:pd_jc}\textbf{c} illustrates in detail the structures of some of the local current configurations for $M=2$. In the FM-phase, as well as in the BF phase, the currents circulate through the boundary of the system, currents along the synthetic direction being strongly suppressed. Interestingly, the FFLO phase (where $c=2$) looks like a Vortex-phase with several vortices.

\clearpage

\setcounter{page}{1}
\setcounter{figure}{0}


\end{document}